\documentclass[12pt]{article}%
\usepackage{amssymb}
\usepackage{amsmath}
\usepackage{sw20elba}
\usepackage[hypertex, bookmarksnumbered=true]{hyperref}
\usepackage{graphicx}
\usepackage{numinsec}
\usepackage{amsfonts}%
\providecommand{\U}[1]{\protect\rule{.1in}{.1in}}

\ifx\pdfoutput\relax\let\pdfoutput=\undefined\fi
\newcount\msipdfoutput
\ifx\pdfoutput\undefined\else
\ifcase\pdfoutput\else
\ifx\paperwidth\undefined\else
\ifdim\paperheight=0pt\relax\else\pdfpageheight\paperheight\fi
\ifdim\paperwidth=0pt\relax\else\pdfpagewidth\paperwidth\fi
\fi\fi\fi
\begin{document}

\author{Alexander Figotin and Guillermo Reyes\\Department of Mathematics\\University of California at Irvine\\Irvine, CA 92697-3875}
\title{Lagrangian Variational Framework for boundary value problems}
\maketitle

\begin{abstract}
A boundary value problem is commonly associated with constraints imposed on a
system at its boundary. We advance here an alternative point of view treating
the system as interacting "boundary" and "interior" subsystems. This view is
implemented through a Lagrangian framework that allows to account for (i) a
variety of forces including dissipative acting at the boundary; (ii) a
multitude of features of interactions between the boundary and the interior
fields when the boundary fields may differ from the boundary limit of the
interior fields; (iii) detailed pictures of the energy distribution and its
flow; (iv) linear and nonlinear effects. We provide a number of elucidating
examples of the structured boundary and its interactions with the system
interior. We also show that the proposed approach covers the well known
boundary value problems.

\end{abstract}

\section{Introduction\label{SectIntro}}

A conventional \emph{boundary value problem} is defined by evolution equations
governing the dynamics of relevant fields at interior points combined with
\emph{boundary conditions} (equations) to form a well-posed problem. Often,
such conditions are imposed \textit{ad hoc} based on physical considerations
at the boundary points. The well-posedness of the resulting boundary value
problem has to be established then independently. The usage of the term
"boundary" is justified by its direct relation to the physical (geometric)
boundary $B=\partial D$ of a spatial domain $D$ associated with the system of
relevant scalar \emph{interior fields}
\[
\psi_{\mathrm{D}}^{\ell}\left(  x,t\right)  ,\qquad x\in D,\quad
\ell=1,2,...k,
\]
defining the system interior state (configuration) at time $t$. A boundary
value problem consists of equations satisfied by $\psi_{\mathrm{D}}^{\ell
}\left(  x,t\right)  $ on $D$, \ as well as \emph{boundary conditions} to be
satisfied by the limit values of $\psi_{\mathrm{D}}^{\ell}\left(  x,t\right)
$ as one approaches the boundary $B$ from the interior. Those conditions may
also involve normal or more general derivatives of the interior fields,
tangential derivatives of the limit field at the boundary, and time
derivatives. There is a variety of boundary conditions used in applications
including fixed boundary, free boundary, non-slip condition, transmission and
impedance conditions and more. Mathematically, fixed boundary and no-slip
conditions correspond to Dirichlet conditions, whereas free boundary
corresponds to Neumann conditions.

We advance in this paper an alternative view on the "boundary value" problem.
A principal difference of our approach from described above conventional one
is that \emph{we expand the space of system states by introducing boundary
fields} $\psi_{\mathrm{B}}^{\ell}(b,t)$, $b\in B$, \emph{as independent
additional degrees of freedom}. Consequently, the states of our system are
described by the pair $\left\{  \psi_{\mathrm{D}}^{\ell},\psi_{\mathrm{B}%
}^{\ell}\right\}  $ of interior and boundary fields . We also acknowledge the
fact that real boundaries are rather thin \textit{interfaces} which can be
thought of as lower dimensional objects, but which usually provide for the
interaction between exterior and interior, and this interaction may be
nontrivial, in particular energy can be stored at the interface. To account
for this fact, we distinguish between the proper boundary fields
$\psi_{\mathrm{B}}^{\ell}$ and the limit values of the interior fields%
\begin{equation}
\psi_{\mathrm{L}}^{\ell}(b,t):=\lim_{x\rightarrow b}\psi_{\mathrm{D}}^{\ell
}(x,t),\qquad b\in B,\quad\ell=1,2,...k \label{psiDpsi1}%
\end{equation}
which are, in general, different from $\psi_{\mathrm{B}}^{\ell}.$ Interactions
between the "two sides" of the boundary are supposed to depend on both
$\psi_{\mathrm{L}}^{\ell}$ and $\psi_{\mathrm{B}}^{\ell}$. The evolution
equations for the interior fields $\psi_{\mathrm{D}}^{\ell}$ over $D$ are
expected to be complemented with the evolution equations for boundary fields
$\psi_{\mathrm{B}}^{\ell}$ over the boundary $B,$ as well as a mathematical
description of the interaction between $\psi_{\mathrm{B}}^{\ell}$ \ and
$\psi_{\mathrm{L}}^{\ell}$. This complementary information constitutes the so
called boundary conditions. The conventional set up can be recovered within
our approach as limiting case of interaction between $\psi_{\mathrm{B}}^{\ell
}$ \ and $\psi_{\mathrm{L}}^{\ell}$ expressed as \emph{continuity or rigidity
constraint}, namely%
\begin{equation}
\psi_{\mathrm{B}}^{\ell}(b,t)\equiv\psi_{\mathrm{L}}^{\ell}(b,t),\qquad b\in
B,\text{ for conventional boundary value problem.} \label{psiDpsi2}%
\end{equation}

Our motivation for pursuing a more general framework for boundary value
problems is that physical systems we are interested in have fields over
boundaries with their own degrees of freedom subjected to different kinds of
forces that often are not accounted for by conventional boundary value
problems. Physical grounds for these additional boundary features are related
to properties of "real" physical boundaries of non-zero thickness that can be
composed of substances different from that of the system interior. We refer to
such more complex boundaries as \emph{structured} and recognize the importance
of their contribution to overall system dynamics by treating them on equal
footing with the system interior. With that in mind we chose to furnish the
system with its physical properties by means of the Lagrangian field framework
for its superiority and flexibility in modeling complex systems subjected to a
variety of forces. Sometimes we use the term boundary loosely to describe not
only the geometric boundary $B$ but boundary fields $\psi_{\mathrm{B}}^{\ell
}(b,t)$ as well.

Though the treatment of the\ system boundary and its interior on equal footing
is a distinct feature of our approach, we acknowledge their difference
regarding dimensionality by integrating it into the system set up. Namely we
suppose that dimensionality $\dim B$ of boundary manifold $B$ is lower than
the dimensionality $\dim D$ of manifold $D$ associated with the interior
subsystem. Most of the time we simply have $\dim B=\dim D-1$. Notice that
geometric boundaries can be very complex and contain manifolds of different
dimensions lower than the dimension $\dim D$. Observe that usage of fields
with densities defined over manifolds of different dimensions side by side is
an established practice in physics. For instance, in Continuum Mechanics
forces with volume densities are complemented by stresses with surface densities.

Leaving the detailed presentation to the following sections we would like to
summarize the main features of our approach to the boundary value problem: (i)
the states of our system are described by the pair $\left\{  \psi_{\mathrm{D}%
}^{\ell},\psi_{\mathrm{B}}^{\ell}\right\}  $ of interior and boundary fields
treated as independent variables; (ii) the fields dynamics is governed by a
system of Lagrangian densities of different dimensionality%
\begin{equation}
\left\{  L_{\mathrm{D}},L_{\mathrm{B}}+L_{\mathrm{INT}}\right\}
\label{lagDB1}%
\end{equation}
with
\begin{equation}
L_{\mathrm{D}}=L_{\mathrm{D}}(t,x,\psi_{\mathrm{D}}^{\ell},D\psi_{\mathrm{D}%
}^{\ell});\qquad D\psi=\left\{  \partial_{\nu}\psi_{\mathrm{D}}^{\ell}%
,\quad\nu=0,1,\ldots n,\quad\ell=1,\ldots,k\right\}  \label{lagDB2}%
\end{equation}
representing the contribution of the interior fields,%
\[
L_{\mathrm{B}}=L_{\mathrm{B}}(t,b,\psi_{\mathrm{B}}^{\ell},D\psi_{\mathrm{B}%
}^{\ell});\qquad D\psi=\left\{  \partial_{\nu}\psi_{\mathrm{B}}^{\ell}%
,\quad\nu=0,1,\ldots n-1,\quad\ell=1,\ldots,k\right\}
\]
that of the boundary fields ($b$ stands for the local coordinates on $B$) and
the interaction Lagrangian
\begin{equation}
L_{\mathrm{INT}}=L_{\mathrm{INT}}(t,b,\psi_{\mathrm{B}}^{\ell},D\psi
_{\mathrm{B}}^{\ell},\psi_{\mathrm{L}}^{\ell}) \label{lagDB3}%
\end{equation}
depending on $\psi_{\mathrm{B}}^{\ell}$ \ and $\psi_{\mathrm{L}}^{\ell}$
defined by (\ref{psiDpsi1}); (iii) external forces acting on our system are
incorporated directly either in the boundary Lagrangian $L_{\mathrm{B}}$ or in
the resulting boundary Euler-Lagrange equations as in the case of frictional
forces. The forces can be of the most general nature, including time
dispersive dissipative forces.

We would like to stress also that the Lagrangian framework involving two
Lagrangian densities as in (\ref{lagDB1})-(\ref{lagDB3}) is of a paramount
importance to our approach for its constructiveness and ability to account for
the energy exchange between the interior and boundary fields $\psi
_{\mathrm{D}}^{\ell}$ and $\psi_{\mathrm{B}}^{\ell}$. The corresponding
evolution equations are (i) the Euler-Lagrange (EL) evolution equations for
the fields inside the region $D$, henceforth referred to as \emph{Domain
Euler-Lagrange equations} (DEL); (ii) the \emph{Interface Euler-Lagrange
equations }(IEL) describing the interior-boundary interaction (between the
fields $\psi_{\mathrm{L}}^{\ell}$ and $\psi_{\mathrm{B}}^{\ell})$ and (iii)
the EL equation for the boundary $B$, henceforth called \emph{Boundary
Euler-Lagrange equation} (BEL). The interaction of the fields $\psi
_{\mathrm{L}}^{\ell}$ on the boundary fields $\psi_{\mathrm{B}}^{\ell}$ is
represented in the IEL equations by "domain" forces, and, similarly, the IEL
equations contain "boundary" forces acting on the interior and adhering to
"action equals reaction" principle. Particular features of those forces are
implemented through the interaction Lagrangian $L_{\mathrm{INT}}$.

The advanced here Lagrangian treatment of "boundary" systems naturally leads
to the consideration of curved Riemannian manifolds. We derive the
Euler-Lagrange equations for Lagrangian systems defined on curved manifolds,
as well as the corresponding energy conservation law. The main difference with
respect to the standard case of a flat manifold is that partial derivatives
are replaced by covariant derivatives.

Observe also that the proposed set up is possible due to the independent
nature of the boundary fields $\psi_{\mathrm{B}}^{\ell}$. Interestingly, even
for conventional boundary value problems this approach, while yielding already
known equations provides a cleaner interpretation of the terms, see subsection
\ref{AdvIndependentBoundary}.

\subsection{Standard variational approach to boundary value problems}

A great deal of research has been conducted to construct and advance boundary
value problems. Mathematical aspects of this research were focused on: (i)
characterization of the special functional spaces that account for a variety
of boundary constraints; (ii) the effect of boundary constraints on the system
spectrum; (iii) an integration into the boundary conditions of frequency
dependent forces, as well as forces of a more general nature; (iv) development
of more flexible variational formulations of the boundary problems. R. Courant
and D. Hilbert consider in their classical book \cite[p. 209]{CH} variational
problems in which the relevant functional depends on "boundary values" of the
function-argument, (see also \cite{FG}). The stationarity principle then
allows to recover both the main partial differential equation obeyed at the
interior, as well as the boundary conditions. Let us briefly recall the
procedure on a simple example.

\begin{figure}
[ptb]
\begin{center}
\ifcase\msipdfoutput
\includegraphics[
height=3.9237in,
width=6.2941in
]%
{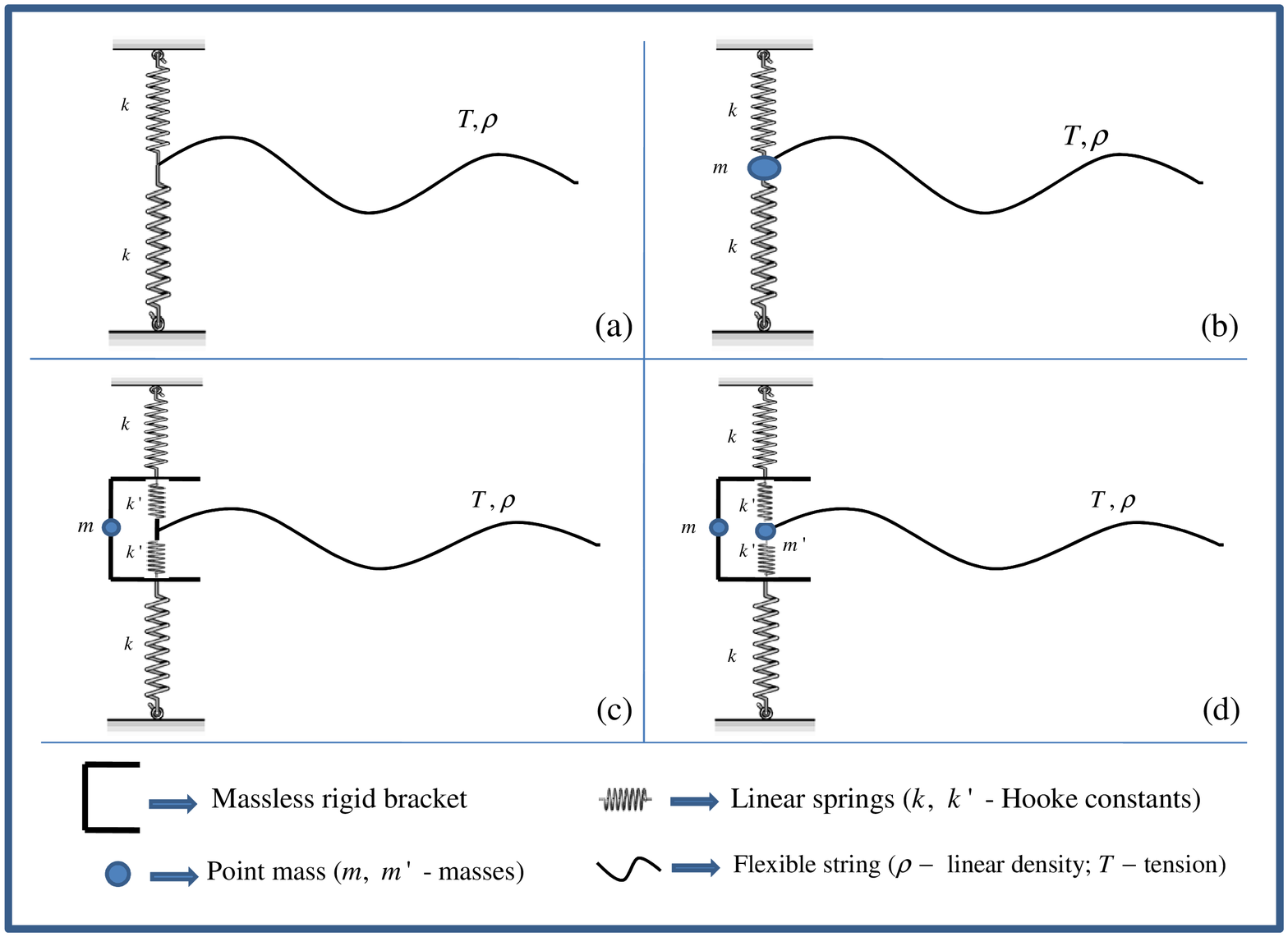}%
\else
\includegraphics[
height=3.9237in,
width=6.2941in
]%
{D:/Initial-BoundaryValueProblem/ArXiv/graphics/FigureFR2Mejor__1.pdf}%
\fi
\caption{Four different structured boundaries for a semi-infinite oscillating
string: a) spring; b) spring+mass; c) and d) more complex spring-mass
systems.}%
\label{Figure1}%
\end{center}
\end{figure}

Consider a transversally oscillating string attached to (massless) spring at
its ends, as in Figure \ref{Figure1} (a), where only one end of the string is
represented. Suppose the string is stretched along the $z$-interval $[0,l]$
and ideal, linear springs are attached at $z=0$ and $z=l$ in such a way that
only transversal oscillations are allowed. If $u=u(z,t)$ stands for the
deflection of the string, the total Lagrangian density is%
\begin{equation}
L\mathcal{(}u_{t},u_{z},u_{0})=\frac{\rho}{2}(u_{t})^{2}-\frac{T}{2}%
(u_{z})^{2}-\frac{k}{2}u_{0}^{2}\delta(z)-\frac{k^{\prime}}{2}u_{l}^{2}%
\delta(z-l), \label{Lrouu1}%
\end{equation}
where $k,k^{\prime}>0$ are the Hooke constants of the springs, $\delta(z)$ is
the Dirac $\delta$-function, $u_{0}=u(0,t)$ and $u_{l}=u(l,t)$. Compactly
supported on $(0,l)$ variations $\delta u$ yield the standard wave equation
for the evolution in the interior,%
\begin{equation}
u_{tt}=a^{2}u_{zz};\qquad a^{2}=\frac{T}{\rho}, \label{Lrouu2}%
\end{equation}
whereas more general variations involving the values at the boundary lead to
the well-known mixed (Robin) boundary conditions%
\begin{equation}
\left(  Tu_{z}-ku\right)  (0,t)=0,\qquad\left(  Tu_{z}+k^{\prime}u\right)
(l,t)=0\qquad t\geq0 \label{BdryCond}%
\end{equation}
that constitute the conditions of equilibrium of forces at $z=0$ and $z=l$. In
the particular case $k=0$ (free string, no spring attached at $z=0$) we get
the Neumann or natural boundary conditions at $z=0$, and the same applies to
$z=l$. The so called Dirichlet boundary condition $u(0,t)=0$, as pointed out
by Courant, \cite{C}, can be seen as a limit case as $k\rightarrow\infty$ (the
same applies to $z=l$). Observe that a \textit{continuity} assumption is
implicit in the derivation: it is assumed that the boundary value of the field
reduces to the limit value of the interior field at the boundary point $z=0$,%
\begin{equation}
u(0,t)=u_{0}(t) \label{Lrouu3}%
\end{equation}
(the same applies to $z=l$). If an additional boundary force is present, it is
added to the boundary condition (\ref{BdryCond}).

The case of a semi-infinite string can be treated analogously. In that case,
however, one must impose some condition at infinity to uniquely determine the
solution. A common choice is the so-called "non-radiation" condition that
excludes perturbations coming from infinity. Such condition will be used on
the one-dimensional examples presented in Subsection
\ref{SubSectOneDimExamples}.

The recent paper by G. Goldstein, \cite{G} deals with the issue of derivation
of general boundary conditions from variational principles. The way the
boundary is incorporated in \cite{G} is based on the consideration of an
extended basic Hilbert space of functions defined on the closure of the
domain\ with a measure supported on the boundary. Such singular measure allows
to incorporate the energy stored in the boundary. All standard boundary
conditions, as well as the less known Wentzell boundary condition, are
recovered in \cite{G} using this construction. Yet another advantage of this
approach is that the obtained EL interior-boundary operator is self-adjoint in
the "mixed" Hilbert space.%

\subsection{Independent boundary fields and their
advantages.\label{AdvIndependentBoundary}}

Our first step towards a more transparent and flexible treatment consists of a
conceptual differentiation between the limit value of the field $u(0,t),$
$u(l,t)$ and \textit{new degrees of freedom}, $u_{0}(t)$ and $u_{l}(t),$
characterizing the boundary subsystem. From this point of view, the
\textit{constraint} $u(0,t)=u_{0}$ is enforced in the previous procedure when
integration by parts is performed. This reduces all possible independent
variations to those for which $\delta u(0,t)=\delta u_{0}.$ The boundary
condition (\ref{BdryCond}) now appears under a different light, as the BEL
equations%
\[
Tu_{z}(0,t)-ku_{0}(t)=0,\qquad Tu_{z}(l,t)+ku_{l}(t)=0\qquad t\geq0.
\]
At first sight this might seem to be a purely formal procedure which does not
give anything new. Observe however that, first of all, the above boundary
conditions yield a clear interpretation: the term $Tu_{z}(0,t)$ represents the
force exerted by the interior upon the boundary $z=0$, while $-ku_{0}(t)$
represents the force exerted by the boundary springs on the ends of the
string. Moreover, if an external force acts on the system through its
boundary, this force can be naturally added to the right-hand side of the BEL
equation, as is customary in the Lagrange formalism. \textit{Without an
explicit distinction of the boundary field, there is no clear guidance on how
to incorporate forces outside the variational setting}. These forces can be of
a very general nature, in particular, dissipative forces with time dispersion.
One of our original motivations to study boundary interaction was precisely to
give a clean variational interpretation of a boundary problem arising in the
modelling of a traveling wave tube (TWT) microwave amplifier. There, complex
impedance conditions on the boundary appear naturally.

Once the boundary and the interior fields are clearly separated we can
introduce general interactions between them through an \emph{interaction
Lagrangian} $L_{\mathrm{INT}}$. For reasons explained below it is natural to
assume that the interaction Lagrangian $L_{\mathrm{INT}}$ depends on the
boundary fields $u_{0},u_{l}$\ and the limit fields $u(0,t),$ $u(l,t)$. The
equalities $u(0,t)=u_{0},$ $u(l,t)=u_{l}$ are not assumed anymore.

Let us consider the case of semi-infinite string, attached to a point
mass-spring system through a secondary spring with the Hooke constant
$\widetilde{k}$, as in Figure \ref{Figure1} (c). In this case our boundary
system is the point mass-spring system, and the interaction with the string is
described by a new term in the Lagrangian
\begin{equation}
L=L_{\mathrm{INT}}=-\frac{\widetilde{k}}{2}(u(0,t)-u_{0})^{2},
\label{IntroEq1}%
\end{equation}
representing the potential energy stored in the secondary spring. The boundary
subsystem, if considered by itself, turns out to be a new and interesting
system. One of its important features is that the effective damping force on
the attached mass is not instantaneous. Indeed, if we assume that the string
was initially at rest and no wave comes from $z=+\infty,$ the evolution of the
position of the mass $u_{0}$ is governed by the integro-differential equation
\begin{equation}
m\partial_{tt}u_{0}+ku_{0}+\widetilde{k}\int_{0}^{t}\mathrm{e}^{-\frac
{a\widetilde{k}}{T}(t-\tau)}\partial_{t}u_{0}(\tau)\,\mathrm{d}\tau=0,
\label{IntroEq}%
\end{equation}
involving a non-local in time friction term (dissipation with dispersion). As
$\widetilde{k}$ increases the secondary springs become more rigid leading in
the limit to a simpler system represented in Figure \ref{Figure1} (b). This
system is known as the Lamb model. In the limit $\widetilde{k}\rightarrow
\infty$ we recover the continuity constraint $u(0,t)=u_{0}$ and the friction
becomes instantaneous, and equation (\ref{IntroEq}) turns into the standard
damped oscillator equation%
\begin{equation}
m\partial_{tt}u_{0}+\frac{T}{a}\partial_{t}u_{0}+ku_{0}=0. \label{IntroEq0}%
\end{equation}
A detailed analysis of the above examples is presented in Subsection
\ref{SubSubLambModified}.

\begin{figure}
[h]
\begin{center}
\ifcase\msipdfoutput
\includegraphics[
trim=0.000000in 1.224454in 0.000000in 0.728378in,
height=3.3001in,
width=6.4947in
]%
{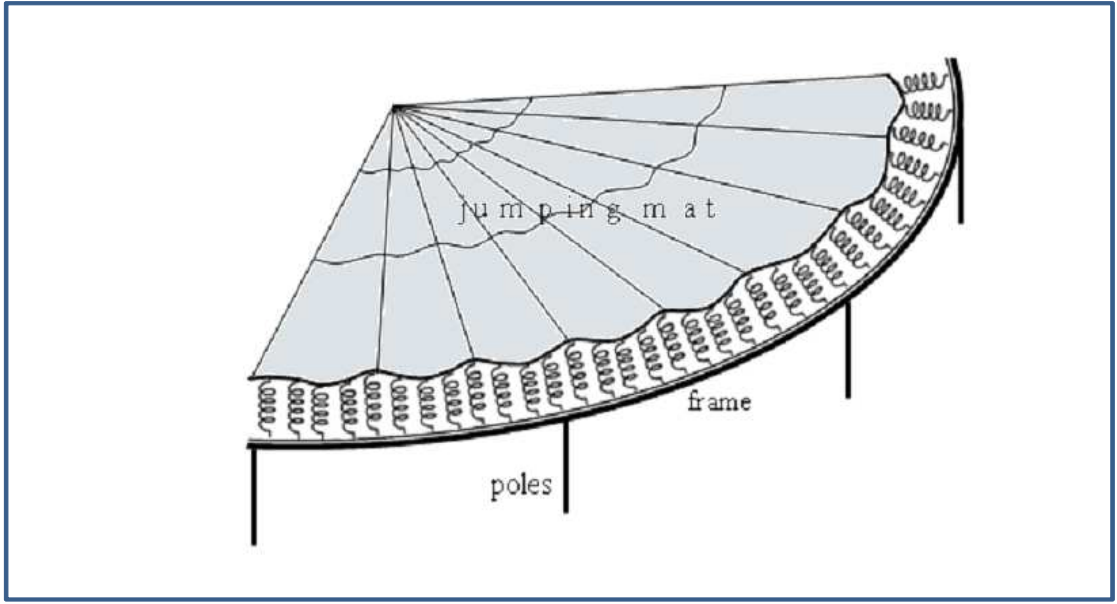}%
\else
\includegraphics[
height=3.3001in,
width=6.4947in
]%
{D:/Initial-BoundaryValueProblem/ArXiv/graphics/Figure3IntentoBIS__2.pdf}%
\fi
\caption{Section of a trampoline.}%
\label{Trampoline}%
\end{center}
\end{figure}

Yet another "real" physical example of a system with a structured boundary is
a trampoline. A trampoline is a device used as a springboard and landing area
in doing acrobatic or gymnastic exercises, see Figure \ref{Trampoline}. It is
made up by an elastic membrane attached by a number of springs to a frame,
typically rectangular or circular. The springs have a twofold effect: that of
providing for a large horizontal tension in the membrane and that of supplying
an elastic support for transversal oscillations along the boundary of the
membrane. The frame is supported by several vertical poles
The trampoline frame is the natural boundary in this example. If we take a
realistic assumption that it is flexible rather than rigid, or that the poles
act as an elastic support, or the both, we obtain a system with a structured
boundary with infinitely many degrees of freedom. We treat this
two-dimensional example in detail in Subsection \ref{SubSectMembrane}.

In fact one can take an even more general approach than advanced here
considering two coupled subsystems constituting a single conservative system.
One of these subsystems of normally lower dimensions can be called the
boundary subsystem and another one of higher dimensions, the interior
subsystem. Particular features of this set up depend on the nature of the
coupling between the two systems. General aspects of linear coupled subsystems
constituting a single conservative system were studied in \cite{FShi1},
\cite{FShi2}.

In the case of a dissipative system when the dissipation can be attributed to
a set of points in the space this set can be treated as a boundary.
Consequently, the proposed here boundary treatment becomes relevant to the
canonical conservative extension of dissipative systems constructed in
\cite{FS}. An example of the kind is a damped oscillator with a retarded
friction function governed by equation (\ref{IntroEq}). The full system
depicted in Figure \ref{Figure1} (c) furnishes a conservative extension
consisting of a string and an additional linear spring.

Comparing our approach to boundary value problems with the one introduced in
\cite{G} we can make the following points : (i) our treatment \ explicitly
separates the boundary from the interior leading to a separate EL equation for
the boundary; (ii) in our approach the interaction between the boundary and
interior fields is introduced by means of an additional interaction Lagrangian
$L_{\mathrm{INT}}$, thus providing for more flexibility in treating
interactions ranging from "no interaction" to a strict continuity constraint.
In the spirit of the classical approach by \cite{CH}, \cite{C}, \cite{FG},
etc. the continuity constraint is assumed in \cite{G}; (iii) our approach
allows to treat general external forces at the boundary outside the
variational framework in a natural way, namely as forces associated to
boundary degrees of freedom.

\subsection{On boundary-interior interactions\label{SubsectBoundInt}}

Though our treatment of boundary value problems allows for a wide range of
boundary-interior interactions, we would like to single out an important class
of interactions accounted for in most of physical situations. This class of
interactions satisfies the following fundamental physical principle.

\begin{itemize}
\item \textsl{Locality principle}: we assume that the boundary interacts with
the interior only through the interior points which are infinitesimally close
to the boundary. A mathematical consequence of this principle is that the
relevant equations are differential in nature, with respect to the spatial variables.
\end{itemize}

Besides the locality principle, we make the following assumptions:

\begin{itemize}
\item \textsl{Action equals reaction (the Third Newton Law)}: we assume that
the force exerted by the boundary on the interior is equal in magnitude and
opposite to the force exerted by the interior on the boundary. These forces
are understood in the generalized sense of the Lagrangian formalism.

\item Let a system be defined on a manifold of dimension $n$ by means of a
first order Lagrangian (depending only on first derivatives). Then, it only
interacts directly with the $(n-1)$-dimensional piece of its boundary manifold.
\end{itemize}

The latter assumption can be illustrated by the following example: suppose we
have a system defined on all of the three-dimensional space except for a
closed, two-dimensional disk. The boundary of this disk is a one-dimensional
circle. We assume that the "bulk" systems interacts with the boundary
subsystem defined on the disk, but does not interact with the independent
field defined on the circle. However, the subsystems defined on the disk and
on its boundary (the circle) do interact.

A rationale for this assumption is as follows: The natural space to deal with
first order Lagrangians is $H^{1}(\Omega),$ the Sobolev space of functions
with first derivative in $L^{2}(\Omega),\dim\Omega=n.$ It is well known that
such functions have well defined traces on smooth boundaries of dimension
$(n-1),$ but not in general on lower dimensional boundaries. Thus, a function
in $H^{1}$ of a three-dimensional ball does not have, in general, a
well-defined trace on its equator. In case of higher order Lagrangians, more
regular spaces are involved and more distant (in the dimensional sense)
interactions are not excluded.

The above assumptions impose certain restrictions in the form of the
interaction terms in the Lagrangian of the system as discussed in Section
\ref{SectLagrangianSetting}.

\subsection{Organization of the paper}

The paper is organized as follows: In Section \ref{SectLagrangianSetting} we
present our general Lagrangian setting for boundary value problems. Subsection
\ \ref{SubsectOneSpaceDim} contains the detailed derivation of the equations
in one spatial dimension, while subsection \ref{SubSectMultipleDimensions}
\ describes the generalization to several space dimensions. The next section
\ref{SectEnergy} discusses the important issue of energy transfer between the
"bulk" system and the boundary, both in one and several spatial dimensions.
Section \ref{SectExamples} is devoted to examples, both in one and two
dimensions. Section \ref{SectConservativeExtensions} discusses how our
approach to boundary interaction is related to the results in \cite{FS} on the
conservative extension of very general dissipative and dispersive systems.

In the final Appendix, we gather some auxiliary material that helps keep the
exposition self-contained. In particular, we briefly recall some facts and
terminology related to dispersive dissipative forces and linear response
theory, as well as some facts from Riemannian differential geometry needed in
the computations. We also include here the derivation of the energy
conservation law, both in the standard case and in the case of a system
defined on a Riemannian manifold, which is a main concern of the paper.

\section{Lagrangian setting for boundary value
problems\label{SectLagrangianSetting}}

\subsection{Problems on one spatial dimension\label{SubsectOneSpaceDim}}

In this section we present a general approach to the formulation of boundary
value problems for the evolution of $k$ scalar fields defined on a
one-dimensional space interval $[b_{1},b_{2}]$.The first subsection is devoted
to fix some notation and to present our final boundary value problem on one
space dimension. The following subsection contains a detailed proof.

\subsubsection{Formulation of the main result in 1D\label{MainResult1D}}

Let us fix some notation: we denote by $\psi^{\ell}(z,t)$ for $t\geq0,$
$b_{1}\leq z\leq b_{2}$, $\ell=1,\ldots,k,$ a set of $k$ scalar fields defined
on a closed interval. As we explained in the Introduction, we allow for a
great flexibility concerning the link between the state at the boundary and
the state in the interior. Thus, we consider separately the restriction of
$\psi^{\ell}$ to the interior of the interval, denoted $\psi_{\mathrm{D}%
}^{\ell},$ and the restriction to the boundary, denoted by $\psi_{\mathrm{B}%
}^{\ell}(b,t)$ and defined on $\left\{  b_{1},b_{2}\right\}  \times
\lbrack0,\infty)$. Suppose that the evolution of $\psi_{\mathrm{D}}^{\ell}$ is
governed by the Lagrangian density%
\begin{equation}
L_{\mathrm{D}}\mathcal{(}t,z,\psi_{\mathrm{D}}^{\ell}\mathfrak{,}%
D\psi_{\mathrm{D}}^{\ell});\qquad D\psi_{\mathrm{D}}^{\ell}=\left\{
\partial_{\nu}\psi_{\mathrm{D}}^{\ell},\quad\nu=0,1,\quad\ell=1,\ldots
,k\right\}  , \label{onedim1}%
\end{equation}
where $\partial_{0}=\partial_{t}$ and $\partial_{1}=\partial_{z},$ while the
evolution of the fields at the boundary is described by the boundary
Lagrangian
\begin{equation}
L_{\mathrm{B}}\mathcal{(}t,z,\psi_{\mathrm{B}}^{\ell}(b_{1},t)\mathfrak{,}%
\psi_{\mathrm{B}}^{\ell}(b_{2},t)\mathfrak{,\partial}_{0}\psi_{\mathrm{B}%
}^{\ell}(b_{1},t)\mathfrak{,\partial}_{0}\psi_{\mathrm{B}}^{\ell}%
(b_{2},t)),\quad\ell=1,\ldots,k. \label{onedim2}%
\end{equation}
Based on the locality principle for interactions from subsection
\ref{SubsectBoundInt}, we assume that the boundary fields $\psi_{\mathrm{B}%
}^{\ell}(b,t)$ interact with the interior only through the interior points
which are infinitesimally close to the boundary. In order to formalize this
idea, we introduce the limit values of the field at the boundary,
$\psi_{\mathrm{L}}^{\ell}(b,t),$ defined as%
\[
\psi_{\mathrm{L}}^{\ell}(b_{1},t):=\lim_{z\rightarrow b_{1}^{+}}%
\psi_{\mathrm{D}}(z,t);\qquad\psi_{\mathrm{L}}^{\ell}(b_{2},t):=\lim
_{z\rightarrow b_{2}^{-}}\psi_{\mathrm{D}}(z,t).
\]
The above definition of the limit values is purely formal, and is understood
in each case depending on the space of admissible fields considered. For our
exposition, we will assume that $\psi_{\mathrm{D}}^{\ell}$ are continuous up
to the boundary of the interval and the above limits are classical ones.

The interior-boundary interaction is described by an interaction Lagrangian.
In order to satisfy the locality assumption we will require that it depend
only on $\psi_{\mathrm{L}}^{\ell}$ and $\psi_{\mathrm{B}}^{\ell},$ with
separated dependence for $z=b_{1}$ and $z=b_{2}.$That is, we assume the
following structure
\begin{equation}
L_{\mathrm{INT,1}}(t,\psi_{\mathrm{L}}^{\ell}(b_{1},t),\psi_{\mathrm{B}}%
^{\ell}(b_{1},t))+L_{\mathrm{INT,2}}(t,\psi_{\mathrm{L}}^{\ell}(b_{2}%
,t),\psi_{\mathrm{B}}^{\ell}(b_{2},t)). \label{interLagrangian}%
\end{equation}
We might also consider interaction Lagrangians depending on the time
derivatives, but refrain to do it for the sake of simplicity in the
exposition. The form (\ref{interLagrangian}) is still too general. \ Indeed,
if we insist on the validity of the third Newton Law for the boundary-interior
interaction, the dependence of $L_{\mathrm{INT}}$ on $\psi_{\mathrm{L}}^{\ell
}$ and $\psi_{\mathrm{B}}^{\ell}$ can not be arbitrary, but only through their
difference $\psi_{\mathrm{B}}^{\ell}-\psi_{\mathrm{L}}^{\ell},$ as in
(\ref{interLagrangian}). Indeed, according to the general Lagrangian
formalism, the respective interactive forces are%
\[
F_{\mathrm{boundary\rightarrow interior}}^{\ell}=\frac{\partial
L_{\mathrm{INT}}}{\partial\psi_{\mathrm{L}}^{\ell}};\qquad
F_{\mathrm{interior\rightarrow boundary}}^{\ell}=\frac{\partial
L_{\mathrm{INT}}}{\partial\psi_{\mathrm{B}}^{\ell}}.
\]
Thus in order to have $F_{\mathrm{boundary\rightarrow interior}}^{\ell
}=-F_{\mathrm{interior\rightarrow boundary}}^{\ell}$ we need%
\[
\frac{\partial L_{\mathrm{INT}}}{\partial\left(  \psi_{\mathrm{L}}^{\ell}%
+\psi_{\mathrm{B}}^{\ell}\right)  }=0,
\]
and therefore $L_{\mathrm{INT}}$ depends only on $\psi_{\mathrm{B}}^{\ell
}-\psi_{\mathrm{L}}^{\ell}$. The final form for the interaction Lagrangian is
thus%
\begin{equation}
L_{\mathrm{INT,1}}(t,(\psi_{\mathrm{B}}^{\ell}-\psi_{\mathrm{L}}^{\ell}%
)(b_{1},t))+L_{\mathrm{INT,2}}(t,(\psi_{\mathrm{B}}^{\ell}-\psi_{\mathrm{L}%
}^{\ell})(b_{2},t)), \label{InterLagrangianFinal}%
\end{equation}
where, with some abuse of notation, we have kept the names of the functions.
In what follows, we denote the gradients of $L_{\mathrm{INT,1}}%
,L_{\mathrm{INT,2}}$ with respect to their field arguments as%
\begin{equation}
\partial_{\psi}L_{\mathrm{INT,1}}=(\partial_{\psi^{\ell}}L_{\mathrm{INT,1}%
}),\qquad\partial_{\psi}L_{\mathrm{INT,2}}=(\partial_{\psi^{\ell}%
}L_{\mathrm{INT,2}}),\qquad\label{Notation2}%
\end{equation}
respectively.

Hence our system is described by a pair of Lagrangians%
\begin{equation}
\left\{  L_{\mathrm{D}},\quad L_{\mathrm{B}}+L_{\mathrm{INT,1}}%
+L_{\mathrm{INT,2}}\right\}  , \label{onedim3}%
\end{equation}
where the first Lagrangian $L_{\mathrm{D}}$ is the density over interval and
the second one $L_{\mathrm{B}}+L_{\mathrm{INT,1}}+L_{\mathrm{INT,2}}$ is that
of a system with a finite number of degrees of freedom. This important and
distinct feature in the Lagrangian treatment is a direct consequence of our
approach that treats the system interior and its boundary on the same footing
and involves interactions of fields defined over manifolds of \ different
dimensions, namely one-dimensional interval and zero-dimensional boundary.

In subsection \ref{DerivationDEL-BELOneDim}, we show that an application of
the Least Action Principle leads to the following \textsl{general boundary
value problem}:%

\begin{gather}
\text{(DEL)\qquad\quad}\partial_{0}\left(  \frac{\partial L_{\mathrm{D}}%
}{\partial\partial_{0}\psi_{\mathrm{D}}^{\ell}}\right)  +\partial_{1}\left(
\frac{\partial L_{\mathrm{D}}}{\partial\partial_{1}\psi_{\mathrm{D}}^{\ell}%
}\right)  -\frac{\partial L_{\mathrm{D}}}{\partial\psi_{\mathrm{D}}^{\ell}%
}=0;\qquad z\in(b_{1},b_{2}),\qquad\qquad\label{DEL}\\
\text{(IEL)\qquad}-\frac{\partial L_{\mathrm{D}}}{\partial\partial_{1}%
\psi_{\mathrm{D}}^{\ell}}(b_{1},t)+\partial_{\psi^{\ell}}L_{\mathrm{INT,1}%
}=0;\qquad\frac{\partial L_{\mathrm{D}}}{\partial\partial_{1}\psi_{\mathrm{D}%
}^{\ell}}(b_{2},t)+\partial_{\psi^{\ell}}L_{\mathrm{INT,2}}=0,\qquad
\label{IEL}\\
\text{(BEL1)\qquad}-\partial_{\psi^{\ell}}L_{\mathrm{INT,1}}+\frac{\partial
L_{\mathrm{B}}}{\partial\psi_{\mathrm{B}}^{\ell}(b_{1},t)}-\partial_{0}\left(
\frac{\partial L_{\mathrm{B}}}{\partial\partial_{0}\psi_{\mathrm{B}}^{\ell
}(b_{1},t)}\right)  =0,\qquad\qquad\qquad\qquad\label{BEL1}\\
\text{(BEL2)\qquad}-\partial_{\psi^{\ell}}L_{\mathrm{INT,2}}+\frac{\partial
L_{\mathrm{B}}}{\partial\psi_{\mathrm{B}}^{\ell}(b_{2},t)}-\partial_{0}\left(
\frac{\partial L_{\mathrm{B}}}{\partial\partial_{0}\psi_{\mathrm{B}}^{\ell
}(b_{2},t)}\right)  =0,\qquad\qquad\qquad\qquad\label{BEL2}%
\end{gather}
for $\ell=1,2,...k$ and $t>0,$ where the derivatives $\partial_{\psi^{\ell}%
}L_{\mathrm{INT,1}},\partial_{\psi^{\ell}}L_{\mathrm{INT,2}}$ are evaluated at
the points%
\[
(t,(\psi_{\mathrm{B}}^{\ell}-\psi_{\mathrm{L}}^{\ell})(b_{1}%
,t)),\ \text{respectively}\ \ (t,(\psi_{\mathrm{B}}^{\ell}-\psi_{\mathrm{L}%
}^{\ell})(b_{2},t)).
\]
The BEL equations are nothing but the Euler-Lagrange motion equations for the
boundaries, under the action of internal potential forces%
\[
\frac{\partial L_{\mathrm{B}}}{\partial\psi_{\mathrm{B}}^{\ell}(b_{i}%
,t)},\qquad i=1,2
\]
and forces due to interaction with the interior, given by $F_{\mathrm{INT,i}%
}^{\ell}=-\partial_{\psi^{\ell}}L_{\mathrm{INT,i}},$ $i=1,2.$ The IEL reflect
the balance between the force, exerted by the interior on the boundary and the
force exerted from the boundary on the interior, via the interaction Lagrangian.

The above equations must be supplemented by suitable initial conditions%
\begin{align}
\psi_{\mathrm{D}}^{\ell}(z,0)  &  =f(z);\qquad\partial_{t}\psi_{\mathrm{D}%
}^{\ell}(z,0)=g(z),\label{InitialConditions}\\
\psi_{\mathrm{B}}^{\ell}(b_{i},0)  &  =c_{i};\qquad\partial_{t}\psi
_{\mathrm{B}}^{\ell}(b_{i},0)=d_{i}.\nonumber
\end{align}

Observe that (\ref{IEL}), (\ref{BEL1}) and (\ref{BEL2}) provide indirect
relations between $\psi_{\mathrm{D}}^{\ell},\partial_{z}\psi_{\mathrm{D}%
}^{\ell}$ and $\partial_{t}\psi_{\mathrm{D}}^{\ell}$ at $b_{1}$ and $b_{2}$.
In order to find independent boundary conditions for $\psi_{\mathrm{D}}^{\ell
}$ one should first find the boundary evolution. In principle, we can proceed
as follows: first, we get rid of the terms containing $L_{\mathrm{INT}}$ in
each pair of boundary conditions, yielding%
\begin{align}
\frac{\partial L_{\mathrm{D}}}{\partial\partial_{1}\psi_{\mathrm{D}}^{\ell}%
}(b_{1},t)+\frac{\partial L_{\mathrm{B}}}{\partial\psi_{\mathrm{B}}^{\ell
}(b_{1},t)}-\partial_{0}\left(  \frac{\partial L_{\mathrm{B}}}{\partial
\partial_{0}\psi_{\mathrm{B}}^{\ell}(b_{1},t)}\right)   &
=0;\label{GenOneDimBC}\\
-\frac{\partial L_{\mathrm{D}}}{\partial\partial_{1}\psi_{\mathrm{D}}^{\ell}%
}(b_{2},t)+\frac{\partial L_{\mathrm{B}}}{\partial\psi_{\mathrm{B}}^{\ell
}(b_{2},t)}-\partial_{0}\left(  \frac{\partial L_{\mathrm{B}}}{\partial\left(
\partial_{0}\psi_{\mathrm{B}}^{\ell}(b_{2},t)\right)  }\right)   &
=0.\nonumber
\end{align}
From these equations and initial conditions for $\psi_{\mathrm{B}}^{\ell}$ in
(\ref{InitialConditions}), we can find the boundary evolution $\psi
_{\mathrm{B}}^{\ell}(b_{i},t)$ depending on the arbitrary forcing functions%
\begin{equation}
-\frac{\partial L_{\mathrm{D}}}{\partial\partial_{1}\psi_{\mathrm{D}}^{\ell}%
}(b_{1},t),\qquad\frac{\partial L_{\mathrm{D}}}{\partial\partial_{1}%
\psi_{\mathrm{D}}^{\ell}}(b_{2},t). \label{ForcingInterior}%
\end{equation}
Then, we plug the values of $\psi_{\mathrm{B}}^{\ell}(b_{i},t)$ in the
(\ref{IEL}) equations, yielding conditions involving only the values of
$\psi_{\mathrm{D}}^{\ell}$ and $\partial_{1}\psi_{\mathrm{D}}^{\ell}$ at the
boundary, which suffice (along with the initial data for $\psi_{\mathrm{D}%
}^{\ell})$ to solve for the interior system. Once we find $\psi_{\mathrm{D}%
}^{\ell}(z,t),$ we know the forcing functions (\ref{ForcingInterior}) and
hence $\psi_{\mathrm{B}}^{\ell}(b_{i},t).$ The described procedure is rather
cumbersome and sometimes can be avoided, as in explicit examples in Subsection
\ref{SubSectOneDimExamples}.

It should be noted that, formally, the case of a rigid (or continuity)
constraint \ $\psi_{\mathrm{L}}^{\ell}(b_{1},t)=\psi_{\mathrm{B}}^{\ell}%
(b_{1},t)$ is not included in the above derivation. A holonomically
constrained Lagrangian systems can be obtained as a limit of unconstrained
systems as the potentials keeping the system close to the given manifold grow
without limit. This fact was already pointed out by Courant in \cite{C}; see
Arnold, \cite{A} or \cite{RU} for a proof. More precisely, if one introduces a
coordinate $q$, measuring the distance to the given manifold in the
configuration space, the dynamics of the constrained system is recovered by
introducing an interaction potential of the form $U=Kq^{2}$ and letting
$K\rightarrow\infty.$ It turns out that $q\rightarrow0$ in such a way that
$U\rightarrow0.$ Thus formally the pair of equations in (\ref{IEL}),
(\ref{BEL1}) and (\ref{BEL2}) dealing with each boundary point is replaced by
the corresponding equation in (\ref{GenOneDimBC}) plus the continuity
constraint. \ The equations in (\ref{GenOneDimBC}) then directly furnish the
boundary conditions for the interior problem. In some of the examples in
section \ref{SubSectOneDimExamples} we carry out this limit process explicitly.

\medskip According to the general Lagrangian formalism, any additional force
acting on the system through its boundary should be added to the left hand
side of the BEL equations, that is, to the left hand side of the second
equations in each of (\ref{BEL1}) and (\ref{BEL2}). Moreover, if the forces
$F_{1}^{\ell}$, $F_{2}^{\ell}$ are potential (monogenic), they can be included
directly in the corresponding boundary Lagrangian. A simple important case is
that in which $F_{1}^{\ell}$ and/or $F_{2}^{\ell}$ do not depend on
$\psi_{\mathrm{B}}^{\ell}(b_{1},t)$ respectively $\psi_{\mathrm{B}}^{\ell
}(b_{2},t).$The associated potential energies in this case are $U_{1}%
(t,\psi_{\mathrm{B}}^{\ell})=-F_{1}^{\ell}\psi_{\mathrm{B}}^{\ell}(b_{1},t),$
$U_{2}(t,\psi_{\mathrm{B}}^{\ell})=-F_{2}^{\ell}\psi_{\mathrm{B}}^{\ell}%
(b_{2},t)$ respectively.

\subsubsection{Derivation of the system of Euler-Lagrange
equations\label{DerivationDEL-BELOneDim}}

In this subsection we provide the details of the derivation of (\ref{DEL}%
)-(\ref{BEL2}) from the Least Action Principle. The action functional
corresponding to the Lagrangian in $L$ in (\ref{onedim3}) is given by%
\begin{equation}
S\left[  \psi_{\mathrm{D}}^{\ell},\psi_{\mathrm{B}}^{\ell}\right]
=\int_{t_{0}}^{t_{1}}\int_{b_{1}}^{b_{2}}L_{\mathrm{D}}\mathrm{dzdt}%
+\int_{t_{0}}^{t_{1}}\left[  L_{\mathrm{B}}+L_{\mathrm{INT,1}}%
+L_{\mathrm{INT,2}}\right]  \,\mathrm{dt.} \label{Action1D}%
\end{equation}
As usual, we start by taking variations $\delta\psi_{\mathrm{D}}^{\ell}$
compactly supported in the space-time domain $(b_{1},b_{2})\times(t_{0}%
,t_{1}),$ while keeping $\psi_{\mathrm{B}}^{\ell}$ fixed, $\delta
\psi_{\mathrm{B}}^{\ell}=0.$ Enforcing $\delta S=0,$ the second integral above
does not contribute to the variation of the action and we arrive at the
equation (\ref{DEL}) satisfied by $\psi_{\mathrm{D}}^{\ell}(z,t)$ in
$(b_{1},b_{2})\times\lbrack0,\infty):$%
\begin{equation}
\partial_{0}\left(  \frac{\partial L_{\mathrm{D}}}{\partial\partial_{0}%
\psi_{\mathrm{D}}^{\ell}}\right)  +\partial_{1}\left(  \frac{\partial
L_{\mathrm{D}}}{\partial\partial_{1}\psi_{\mathrm{D}}^{\ell}}\right)
-\frac{\partial L_{\mathrm{D}}}{\partial\psi_{\mathrm{D}}^{\ell}}=0.
\label{ELEq}%
\end{equation}
Generically, (\ref{ELEq}) is a system of $k$ scalar, second order in $z$ and
$t$ partial differential equations.

Next we take general, independent variations of both the interior and the
boundary fields, assuming only that%
\[
\delta\psi_{\mathrm{D}}^{\ell}(z,t_{0})=\delta\psi_{\mathrm{D}}^{\ell}%
(z,t_{1})=\delta\psi_{\mathrm{B}}^{\ell}(t_{0})=\delta\psi_{\mathrm{B}}^{\ell
}(t_{1})=0.
\]
The variation of the action is%
\begin{gather}
\delta S\left[  \psi_{\mathrm{D}}^{\ell},\psi_{\mathrm{B}}^{\ell}\right]
(\delta\psi_{\mathrm{D}}^{\ell},\delta\psi_{\mathrm{B}}^{\ell}%
)=\label{VarAction1D}\\
=\int_{t_{0}}^{t_{1}}\int_{b_{1}}^{b_{2}}%
{\displaystyle\sum\nolimits_{\ell}}
\left[  \frac{\partial L_{\mathrm{D}}}{\partial\psi_{\mathrm{D}}^{\ell}%
}-\partial_{0}\left(  \frac{\partial L_{\mathrm{D}}}{\partial\partial_{0}%
\psi_{\mathrm{D}}^{\ell}}\right)  -\partial_{1}\left(  \frac{\partial
L_{\mathrm{D}}}{\partial\partial_{1}\psi_{\mathrm{D}}^{\ell}}\right)  \right]
\delta\psi_{\mathrm{D}}^{\ell}(z,t)\,\mathrm{dzdt}+\nonumber\\
+\int_{t_{0}}^{t_{1}}%
{\displaystyle\sum\nolimits_{\ell}}
\left[  \frac{\partial L_{\mathrm{D}}}{\partial\partial_{1}\psi_{\mathrm{D}%
}^{\ell}}(b_{2},t)\delta\psi_{\mathrm{L}}^{\ell}\,(b_{2},t)-\frac{\partial
L_{\mathrm{D}}}{\partial\partial_{1}\psi_{\mathrm{D}}^{\ell}}(b_{1}%
,t)\delta\psi_{\mathrm{L}}^{\ell}(b_{1},t)\right]  \,\mathrm{dt}\,+\nonumber\\
+\int_{t_{0}}^{t_{1}}%
{\displaystyle\sum\nolimits_{\ell}}
\left[  \partial_{\psi^{l}}L_{\mathrm{INT,1}}(\delta\psi_{\mathrm{L}}^{\ell
}-\delta\psi_{\mathrm{B}}^{\ell})(b_{1},t)+\partial_{\psi^{l}}%
L_{\mathrm{INT,2}}(\delta\psi_{\mathrm{L}}^{\ell}-\delta\psi_{\mathrm{B}%
}^{\ell})(b_{2},t)\right]  \,\mathrm{dt+}\nonumber\\
+\int_{t_{0}}^{t_{1}}%
{\displaystyle\sum\nolimits_{\ell}}
\left[  \frac{\partial L_{\mathrm{B}}}{\partial\psi_{\mathrm{B}}^{\ell}%
(b_{2},t)}-\partial_{0}\left(  \frac{\partial L_{\mathrm{B}}}{\partial
\partial_{0}\psi_{\mathrm{B}}^{\ell}(b_{2},t)}\right)  \right]  \delta
\psi_{\mathrm{B}}^{\ell}(b_{2},t)\,\mathrm{dt+}\nonumber\\
+\int_{t_{0}}^{t_{1}}%
{\displaystyle\sum\nolimits_{\ell}}
\left[  \frac{\partial L_{\mathrm{B}}}{\partial\psi_{\mathrm{B}}^{\ell}%
(b_{1},t)}-\partial_{0}\left(  \frac{\partial L_{\mathrm{B}}}{\partial
\partial_{0}\psi_{\mathrm{B}}^{\ell}(b_{1},t)}\right)  \right]  \delta
\psi_{\mathrm{B}}^{\ell}(b_{1},t)\,\mathrm{dt,}\nonumber
\end{gather}
where we put, for simplicity,%
\[
\frac{\partial L_{\mathrm{D}}}{\partial\partial_{1}\psi_{\mathrm{D}}^{\ell}%
}(b_{2},t):=\lim_{z\rightarrow b_{2}^{-}}\frac{\partial L_{\mathrm{D}}%
}{\partial\partial_{1}\psi_{\mathrm{D}}^{\ell}}\mathcal{(}t,z,\psi
_{\mathrm{D}}^{\ell}(z,t)\mathfrak{,\partial}_{0}\psi_{\mathrm{D}}^{\ell
}(z,t),\partial_{1}\psi_{\mathrm{D}}^{\ell}(z,t)),
\]
and analogously with $\frac{\partial L_{\mathrm{D}}}{\partial\left(
\partial_{1}\psi_{\mathrm{D}}^{\ell}\right)  }(b_{1},t).$ The derivatives
$\partial_{\psi^{l}}L_{\mathrm{INT,1}},$ $\partial_{\psi^{l}}L_{\mathrm{INT,2}%
}$ are evaluated at the point $(t,\psi_{\mathrm{L}}^{\ell}(b_{1}%
,t)-\psi_{\mathrm{B}}^{\ell}(b_{1},t))$, respectively $(t,\psi_{\mathrm{L}%
}^{\ell}(b_{2},t)-\psi_{\mathrm{B}}^{\ell}(b_{2},t))$. We have assumed enough
regularity such that integration by parts in time and space is allowed. The
double integral above vanishes, thanks to (\ref{ELEq}). \ Conveniently
regrouping the terms in the above expression, we get
\begin{gather*}
\delta S=\int_{t_{0}}^{t_{1}}%
{\displaystyle\sum\nolimits_{\ell}}
\left[  \frac{\partial L_{\mathrm{D}}}{\partial\partial_{1}\psi_{\mathrm{D}%
}^{\ell}}(b_{2},t)+\partial_{\psi^{l}}L_{\mathrm{INT,2}}\right]  \delta
\psi_{\mathrm{L}}^{\ell}(b_{2},t)\,\mathrm{dt}\\
+\int_{t_{0}}^{t_{1}}%
{\displaystyle\sum\nolimits_{\ell}}
\left[  -\frac{\partial L_{\mathrm{D}}}{\partial\partial_{1}\psi_{\mathrm{D}%
}^{\ell}}(b_{1},t)+\partial_{\psi^{l}}L_{\mathrm{INT,1}}\right]  \delta
\psi_{\mathrm{L}}^{\ell}(b_{1},t)\,\mathrm{dt}\\
+\int_{t_{0}}^{t_{1}}%
{\displaystyle\sum\nolimits_{\ell}}
\left[  -\partial_{\psi^{l}}L_{\mathrm{INT,2}}+\frac{\partial L_{\mathrm{B}}%
}{\partial\psi_{\mathrm{B}}^{\ell}(b_{2},t)}-\partial_{0}\left(
\frac{\partial L_{\mathrm{B}}}{\partial\partial_{0}\psi_{\mathrm{B}}^{\ell
}(b_{2},t)}\right)  \right]  \delta\psi_{\mathrm{B}}^{\ell}(b_{2}%
,t)\,\mathrm{dt}\\
\mathrm{+}\int_{t_{0}}^{t_{1}}%
{\displaystyle\sum\nolimits_{\ell}}
\left[  -\partial_{\psi^{l}}L_{\mathrm{INT,1}}+\frac{\partial L_{\mathrm{B}}%
}{\partial\psi_{\mathrm{B}}^{\ell}(b_{1},t)}-\partial_{0}\left(
\frac{\partial L_{\mathrm{B}}}{\partial\partial_{0}\psi_{\mathrm{B}}^{\ell
}(b_{1},t)}\right)  \right]  \delta\psi_{\mathrm{B}}^{\ell}(b_{1}%
,t)\,\mathrm{dt}.
\end{gather*}
Arbitrariness and independence of the variations $\delta\psi_{\mathrm{L}%
}^{\ell}$ and $\delta\psi_{\mathrm{B}}^{\ell}$ for $\ell=1,2,...k$ imply the
boundary conditions%
\begin{align}
\frac{\partial L_{\mathrm{D}}}{\partial\partial_{1}\psi_{\mathrm{D}}^{\ell}%
}(b_{2},t)+\partial_{\psi^{l}}L_{\mathrm{INT,2}}  &  =0;\label{4BdryCond}\\
-\frac{\partial L_{\mathrm{D}}}{\partial\partial_{1}\psi_{\mathrm{D}}^{\ell}%
}(b_{1},t)+\partial_{\psi^{l}}L_{\mathrm{INT,1}}  &  =0;\label{4BdryCond2}\\
-\partial_{\psi^{l}}L_{\mathrm{INT,2}}+\frac{\partial L_{\mathrm{B}}}%
{\partial\psi_{\mathrm{B}}^{\ell}(b_{2},t)}-\partial_{0}\left(  \frac{\partial
L_{\mathrm{B}}}{\partial\partial_{0}\psi_{\mathrm{B}}^{\ell}(b_{2},t)}\right)
&  =0;\label{4BdryCond3}\\
-\partial_{\psi^{l}}L_{\mathrm{INT,1}}+\frac{\partial L_{\mathrm{B}}}%
{\partial\psi_{\mathrm{B}}^{\ell}(b_{1},t)}-\partial_{0}\left(  \frac{\partial
L_{\mathrm{B}}}{\partial\partial_{0}\psi_{\mathrm{B}}^{\ell}(b_{1},t)}\right)
&  =0, \label{4BdryCond4}%
\end{align}
constituting (\ref{BEL1})-(\ref{BEL2}).

\subsection{Multidimensional case\label{SubSectMultipleDimensions}}

The above approach can be generalized to several space dimensions. Throughout
this section, $\Omega$ will be an open region of the Euclidean space
$\mathbb{E}^{n}$. The (topological) boundary $\Gamma=\partial\Omega$ of such a
region can be very complicated, but we restrict to the case when it is made up
of a finite number of lower dimensional smooth manifolds. Actually, we
restrict at first to the case in which $\Gamma$ is a closed, smooth
$(n-1)$-dimensional manifold$.$ In particular, this assumption implies that
$\Gamma$ has empty boundary (as a manifold). In subsection \ref{LowerDimBound}
we briefly discuss the modifications needed in case of regions that exhibit
lower dimensional pieces of boundary.

We consider $\Gamma$ as an embedded hypersurface in $\mathbb{E}^{n},$ with
Riemannian structure induced from that of the ambient space. In the sequel, by
$b=(b_{1},b_{2},...b_{n-1})$ we denote local coordinates on $\Gamma$ (strictly
speaking, on one of the charts) and by $g_{\alpha\beta}=g_{\alpha\beta}(b)$
the induced metrics. Finally, we put $g:=\det\left(  g_{\alpha\beta}\right)
.$\textrm{ }

Of concern is the evolution of $k$ scalar fields defined on the closure
$\overline{\Omega}=\Omega+\Gamma:$
\[
\psi^{\ell}:\overline{\Omega}\times\lbrack0,\infty)\rightarrow%
\mathbb{R}
,\qquad\ell=1,2,...k
\]
We call, as above, $\psi_{\mathrm{D}}^{\ell}$ the restriction of $\psi^{\ell}$
to the interior of the region, and $\psi_{\mathrm{B}}^{\ell}$ the field on
$\Gamma.$ We do not impose any \textit{a priori} relation between
$\psi_{\mathrm{D}}^{\ell}$ and $\psi_{\mathrm{B}}^{\ell}.$ We assume, however,
that both fields are smooth enough, in particular that $\psi_{\mathrm{D}%
}^{\ell}$ and its first derivatives admit boundary values and integration by
parts is allowed. Let $x=x(b),$ $b\in D\subset%
\mathbb{R}
^{n-1}$ be a parametrization of $\Gamma$ (strictly speaking, of a chart of
$\Gamma).$Via this parametrization, $\psi_{\mathrm{B}}^{\ell}$ is a function
of the local coordinates chosen in $\Gamma$ and time, $\psi_{\mathrm{B}}%
^{\ell}=\psi_{\mathrm{B}}^{\ell}(b,t)$. As before, we formally define the
limit field on the boundary
\[
\psi_{\mathrm{L}}^{\ell}(b,t):=\lim_{y\rightarrow x(b),\ y\in\Omega}%
\psi_{\mathrm{D}}^{\ell}(y,t)
\]

\subsubsection{Formulation of the main result in multiple dimensions}

Suppose that the dynamics in $\Omega$ is governed by a Lagrangian density%
\[
L_{\mathrm{D}}(t,x,\psi_{\mathrm{D}}^{\ell},D\psi_{\mathrm{D}}^{\ell});\quad
D\psi_{\mathrm{D}}^{\ell}=\left\{  \partial_{\nu}\psi_{\mathrm{D}}^{\ell
},\quad\nu=0,1,2,...n,\quad\ell=1,\ldots,k\right\}  ,\quad x\in\Omega,\quad
t\in\lbrack0,\infty),
\]
where $\partial_{0}=\partial_{t}$, $\partial_{i}=\partial_{x_{i}}$ and
$x_{i},$ $i=1...n,$ are Cartesian coordinates. Much as in the one-dimensional
case, we define a boundary $(n-1)$-dimensional Lagrangian density on $\Gamma,$%
\[
L_{\mathrm{B}}(t,b,\psi_{\mathrm{B}}^{\ell},D\psi_{\mathrm{B}}^{\ell}),\qquad
D\psi_{\mathrm{B}}^{\ell}=\left\{  \partial_{\nu}\psi_{\mathrm{B}}^{\ell
},\quad\nu=0,1,2,...n-1,\quad\ell=1,\ldots,k\right\}  ,
\]
where $\partial_{0}=\partial_{t}$, $\partial_{i}=\partial_{b_{i}}$ and
$b_{i},$ $i=1...n-1,$ are local coordinates. We assume that the interaction
between the "bulk" and the boundary is given by a $(n-1)$-dimensional
Lagrangian density of the form%
\[
L_{\mathrm{INT}}(t,b,\psi_{\mathrm{L}}^{\ell}-\psi_{\mathrm{B}}^{\ell}).
\]
The form of the dependence of $L_{\mathrm{INT}}$ on $\psi_{\mathrm{L}}^{\ell}$
and $\psi_{\mathrm{B}}^{\ell}$ can be justified as in the one-dimensional case
in Subsection \ref{SubsectOneSpaceDim}. We might consider more general
interaction Lagrangians, including dependence on time or tangential
derivatives on the boundary. Such dependence would not violate the third
Newton's Law, but we refrain from considering such general situation for the
sake of simplicity in the exposition. We adhere to notation (\ref{Notation2}),
that is,%
\[
\partial_{\psi^{\ell}}L_{\mathrm{INT}}:=\frac{\partial L_{\mathrm{INT}}%
}{\partial\left(  \psi_{\mathrm{L}}^{\ell}-\psi_{\mathrm{B}}^{\ell}\right)
},\qquad\ell=1,2,...k.
\]
The total Lagrangian corresponding to the densities $\left\{  L_{\mathrm{D}%
},L_{\mathrm{B}}+L_{\mathrm{INT}}\right\}  $ is
\begin{equation}
L=\int_{\Omega}L_{\mathrm{D}}\,\mathrm{dx+}\int_{\Gamma}\left[  L_{\mathrm{B}%
}\mathrm{+}L_{\mathrm{INT}}\right]  \,\mathrm{dS}_{\mathrm{b}}\mathrm{,}
\label{MultiLagrangian}%
\end{equation}
where%
\[
\mathrm{dS}_{\mathrm{b}}=\sqrt{g}\,\mathrm{db}_{1}\mathrm{\wedge db}%
_{2}\mathrm{\wedge...\wedge db}_{n-1},\qquad g:=\det\left(  g_{\alpha\beta
}\right)
\]
is the $(n-1)$-dimensional volume element in local coordinates $(b_{i})$. We
introduce the following notation:%
\begin{gather}
G_{\mathrm{D}}^{\ell0}=\frac{\partial L_{\mathrm{D}}}{\partial\partial_{0}%
\psi_{\mathrm{D}}^{\ell}},\quad G_{\mathrm{D}}^{\ell\nu}=\frac{\partial
L_{\mathrm{D}}}{\partial\partial_{\nu}\psi_{\mathrm{D}}^{\ell}},\quad
G_{\mathrm{D}}^{\ell}=(G_{\mathrm{D}}^{\ell\nu});\qquad\nu=1,\ldots
,n,\quad\ell=1,\ldots,k,\label{Gs}\\
G_{\mathrm{B}}^{\ell0}=\frac{\partial L_{\mathrm{B}}}{\partial\partial_{0}%
\psi_{\mathrm{B}}^{\ell}},\quad G_{\mathrm{D}}^{\ell\nu}=\frac{\partial
L_{\mathrm{B}}}{\partial\partial_{\nu}\psi_{\mathrm{B}}^{\ell}},\quad
G_{\mathrm{B}}^{\ell}=(G_{\mathrm{B}}^{\ell\nu});\qquad\nu=1,\ldots
,n-1,\quad\ell=1,\ldots,k.\nonumber
\end{gather}
Our main result reads as follows: the evolution of the above system is
described by the following \textsl{general} \textsl{boundary value problem in
multiple dimensions:}%
\begin{gather}
\text{(DEL)}\qquad\qquad\qquad\frac{\partial L_{\mathrm{D}}}{\partial
\psi_{\mathrm{D}}^{\ell}}-\operatorname{div}G_{\mathrm{D}}^{\ell}-\partial
_{0}G_{\mathrm{D}}^{\ell0}=0;\qquad x\in\Omega,\quad t>0;\label{MDEL}\\
\text{(IEL)}\qquad\qquad\left\langle G_{\mathrm{D}}^{\ell}%
(x(b),t),\,\mathrm{n}(b)\right\rangle +\partial_{\psi^{\ell}}L_{\mathrm{INT}%
}=0;\qquad b\in\Gamma,\quad t>0;\label{MIEL}\\
\text{(BEL)}\qquad\frac{\partial L_{\mathrm{B}}}{\partial\psi_{\mathrm{B}%
}^{\ell}}-\widetilde{\operatorname{div}}G_{\mathrm{B}}^{\ell}-\partial
_{0}G_{\mathrm{B}}^{\ell0}-\partial_{\psi^{\ell}}L_{\mathrm{INT}}=0;\qquad
b\in\Gamma,\quad t>0, \label{MBEL}%
\end{gather}
for $\ \ell=1,2,...k$. In DEL, $\operatorname{div}$ stands for the standard
divergence, whereas in BEL, $\widetilde{\operatorname{div}}$ stands for the
covariant divergence on $\Gamma$, see formula (\ref{divergenceRiemannian}),
$\mathrm{n=n}(b)$ is the unit normal vector to $\Gamma$. The derivatives
$\partial_{\psi^{j}}L_{\mathrm{INT}}$ are evaluated at the point
$(t,(\psi_{\mathrm{L}}^{\ell}-\psi_{\mathrm{B}}^{\ell})(b,t))$.

Observe the analogy with the one-dimensional system (\ref{DEL})-(\ref{BEL2}).
The general comments made for the one-dimensional case apply without change. A
major difference between the present case and the one-dimensional one
discussed above is the role played by the geometry of the boundary. Indeed,
the spatial variation of $\ g=\det\left(  g_{\alpha\beta}\right)  $ enters
explicitly in the motion equation (\ref{MBEL}) for the boundary system.

External forces applied on the boundary can be added, as usual, to the left
hand side of (\ref{MBEL}).

A general account on Lagrangian formalism on manifolds can be found in
\cite[Section 3]{CaCha}. Since we are dealing with two manifolds of different
dimensions and the interaction is a major issue, we provide an independent
derivation of the E-L equations in the next subsection.

\subsubsection{Derivation of covariant form of the Euler-Lagrange
equations\label{DeriMulti}}

The action \ associated to the Lagrangian (\ref{MultiLagrangian}) from
$t_{0\text{ }}$ to $t_{1}$ is%
\[
S\left[  \psi_{\mathrm{D}}^{\ell},\psi_{\mathrm{B}}^{\ell}\right]
\mathcal{=}\int_{t_{0}}^{t_{1}}\int_{\Omega}L_{\mathrm{D}}\,\mathrm{dxdt}%
+\int_{t_{0}}^{t_{1}}\int_{\Gamma}\left[  L_{\mathrm{B}}\mathrm{+}%
L_{\mathrm{INT}}\right]  \,\mathrm{dS}_{\mathrm{b}}\mathrm{dt.}%
\]
If we take variations $\delta\psi_{\mathrm{D}}^{\ell}$ compactly supported in
$\Omega\times\left(  t_{0,}t_{1}\right)  ,$ while keeping $\psi_{\mathrm{B}%
}^{\ell}$ fixed, the stationarity of $S\left[  \psi_{\mathrm{D}}^{\ell}%
,\psi_{\mathrm{B}}^{\ell}\right]  $ leads to the usual equations (\ref{MDEL})
for the evolution of $\psi_{\mathrm{D}}^{\ell}(x,t)$ \ in $\Omega\times
\lbrack0,\infty),$
\begin{equation}
\frac{\partial L_{\mathrm{D}}}{\partial\psi_{\mathrm{D}}^{\ell}}%
-\operatorname{div}G_{\mathrm{D}}^{\ell}-\partial_{0}G_{\mathrm{D}}^{\ell0}=0.
\label{MultiInterior}%
\end{equation}
Taking into account the above DEL equations, general variations of the fields
satisfying $\delta\psi_{\mathrm{B}}^{\ell}(b,t_{0})=\delta\psi_{\mathrm{B}%
}^{\ell}(b,t_{1})$ yield the following expression for the variation of the
action%
\begin{gather}
\delta S\left[  \psi_{\mathrm{D}}^{\ell},\psi_{\mathrm{B}}^{\ell}\right]
(\delta\psi_{\mathrm{D}}^{\ell},\delta\psi_{\mathrm{B}}^{\ell})=\int_{t_{0}%
}^{t_{1}}\int_{\Omega}%
{\displaystyle\sum\nolimits_{\ell}}
{\displaystyle\sum\limits_{i=1}^{n}}
\partial_{i}\left(  \frac{\partial L_{\mathrm{D}}}{\partial\partial_{i}%
\psi_{\mathrm{D}}^{\ell}}\delta\psi_{\mathrm{D}}^{\ell}\right)
\,\mathrm{dxdt}+\label{ActionMulti}\\
+\int_{t_{0}}^{t_{1}}\int_{\Gamma}%
{\displaystyle\sum\nolimits_{\ell}}
\left[  \frac{\partial L_{\mathrm{B}}}{\partial\psi_{\mathrm{B}}^{\ell}}%
\delta\psi_{\mathrm{B}}^{\ell}+%
{\displaystyle\sum\limits_{j=1}^{n-1}}
\frac{\partial L_{\mathrm{B}}}{\partial\partial_{j}\psi_{\mathrm{B}}^{\ell}%
}\partial_{j}\delta\psi_{\mathrm{B}}^{\ell}-\partial_{0}G_{\mathrm{B}}^{\ell
0}\delta\psi_{\mathrm{B}}^{\ell}\right]  \,\mathrm{dS}_{\mathrm{b}}%
\mathrm{dt}\nonumber\\
+\int_{t_{0}}^{t_{1}}\int_{\Gamma}\left[
{\displaystyle\sum\nolimits_{\ell}}
\partial_{\psi^{\ell}}L_{\mathrm{INT}}(t,(\psi_{\mathrm{L}}^{\ell}%
-\psi_{\mathrm{B}}^{\ell})(b,t))(\delta\psi_{\mathrm{L}}^{\ell}-\delta
\psi_{\mathrm{B}}^{\ell})(b,t)\,\right]  \,\mathrm{dS}_{\mathrm{b}%
}\mathrm{dt,}\nonumber
\end{gather}
Observe that, for each $\ell=1,2,...k$, $G_{\mathrm{D}}^{\ell}$ in (\ref{Gs})
is a genuine (contravariant) vector in $\Omega$. It is easy to check that the
coordinates of $G_{\mathrm{D}}^{\ell}$ transform according to the
contravariant vector law. Indeed, consider any other (in general, curvilinear)
coordinate system $(x_{i}^{\prime})$. \ Since the Lagrangian is a scalar, that
is%
\[
L_{\mathrm{D}}^{\prime}(t,x_{i}^{\prime},D^{\prime}\psi_{\mathrm{D}}^{\ell
})=L_{\mathrm{D}}(t,x_{i},D\psi_{\mathrm{D}}^{\ell}),
\]
and since $\partial_{x_{i}}\psi_{\mathrm{D}}^{\ell}=\sum\partial
_{x_{s}^{\prime}}\psi_{\mathrm{D}}^{\ell}\partial x_{s}^{\prime}/\partial
x_{i}$ we have
\[
G_{\mathrm{D}}^{\ell\nu\prime}=\frac{\partial L_{\mathrm{D}}}{\partial
\partial_{\nu^{\prime}}\psi_{\mathrm{D}}^{\ell}}=\sum_{\varkappa=1}^{n}%
\frac{\partial L_{\mathrm{D}}}{\partial\partial_{\varkappa}\psi_{\mathrm{D}%
}^{\ell}}\frac{\partial x_{\nu}^{\prime}}{\partial x_{\varkappa}}%
=\sum_{\varkappa=1}^{n}\frac{\partial x_{\nu}^{\prime}}{\partial x_{\varkappa
}}G_{\mathrm{D}}^{\ell\varkappa}%
\]
implying that $G_{\mathrm{D}}^{\ell}$ transforms as a vector. Consequently,
the linear combination%
\[
G_{\mathrm{D}}:=%
{\displaystyle\sum\nolimits_{\ell}}
G_{\mathrm{D}}^{\ell}\delta\psi_{\mathrm{D}}^{\ell}%
\]
with scalar fields $\delta\psi_{\mathrm{D}}^{\ell}$ is also a vector. The
standard Divergence (Gauss) Theorem yields
\begin{gather*}
\int_{t_{0}}^{t_{1}}\int_{\Omega}%
{\displaystyle\sum\limits_{i=1}^{n}}
\partial_{i}\left(
{\displaystyle\sum\nolimits_{\ell}}
G_{\mathrm{D}}^{\ell}\delta\psi_{\mathrm{D}}^{\ell}\right)  \,\mathrm{dxdt=}%
\int_{t_{0}}^{t_{1}}\int_{\Omega}\operatorname{div}G_{\mathrm{D}%
}\,\mathrm{dxdt=}\\
=\int_{t_{0}}^{t_{1}}\int_{\Gamma}\left\langle G_{\mathrm{D}}%
(x(b),t)\,,\mathrm{n}(b)\right\rangle \,\mathrm{dS}_{\mathrm{b}}%
\mathrm{dt}=\int_{t_{0}}^{t_{1}}\int_{\Gamma}%
{\displaystyle\sum\nolimits_{\ell}}
\left\langle G_{\mathrm{D}}^{\ell}(x(b),t)\,,\mathrm{n}(b)\right\rangle
\,\delta\psi_{\mathrm{L}}^{\ell}\,\mathrm{dS}_{\mathrm{b}}\mathrm{dt,}%
\end{gather*}
where $\mathrm{n}(b)$ stands for the unit (outward) normal vector to $\Gamma$
and $\left\langle \,,\right\rangle $ stands for the scalar product in
$\mathbb{E}^{n}$.

Next, we integrate by parts the term%
\[
\int_{t_{0}}^{t_{1}}\int_{\Gamma}%
{\displaystyle\sum\nolimits_{\ell}}
{\displaystyle\sum\limits_{j=1}^{n-1}}
\frac{\partial L_{\mathrm{B}}}{\partial\partial_{j}\psi_{\mathrm{B}}^{\ell}%
}\partial_{j}\delta\psi_{\mathrm{B}}^{\ell}\,\mathrm{dS}_{\mathrm{b}%
}\mathrm{dt}%
\]
in (\ref{ActionMulti}). As before, $G_{\mathrm{B}}^{\ell}$ in (\ref{Gs}) are
legitimate vector fields on $\Gamma,$ as well as the combination
\[
G_{\mathrm{B}}:=%
{\displaystyle\sum\nolimits_{\ell}}
G_{\mathrm{B}}^{\ell}\delta\psi_{\mathrm{B}}^{\ell}%
\]
Since we are integrating on a curved manifold, we use \textsl{covariant}
differentiation, see (\ref{covder}) yielding the two integrals%
\begin{gather}
\int_{t_{0}}^{t_{1}}\int_{\Gamma}%
{\displaystyle\sum\nolimits_{\ell}}
\left\{
{\displaystyle\sum\limits_{i=1}^{n-1}}
\widetilde{\partial}_{i}\left(  \frac{\partial L_{\mathrm{B}}}{\partial
\partial_{b_{i}}\psi_{\mathrm{B}}^{\ell}}\delta\psi_{\mathrm{B}}^{\ell
}\right)  \,-%
{\displaystyle\sum\limits_{i=1}^{n-1}}
\widetilde{\partial}_{i}\left(  \frac{\partial L_{\mathrm{B}}}{\partial
\partial_{i}\psi_{\mathrm{B}}^{\ell}}\right)  \delta\psi_{\mathrm{B}}^{\ell
}\right\}  \,\mathrm{dS}_{\mathrm{b}}\mathrm{dt=}\label{IntParts}\\
=\int_{t_{0}}^{t_{1}}\int_{\Gamma}\widetilde{\operatorname{div}}G_{\mathrm{B}%
}\mathrm{dS}_{\mathrm{b}}\mathrm{dt}\,-\int_{t_{0}}^{t_{1}}\int_{\Gamma}%
{\displaystyle\sum\nolimits_{\ell}}
\widetilde{\operatorname{div}}G_{\mathrm{B}}^{\ell}\,\delta\psi_{\mathrm{B}%
}^{\ell}\,\mathrm{dS}_{\mathrm{b}}\mathrm{dt,}\nonumber
\end{gather}
where $\widetilde{\partial}_{i}=\widetilde{\partial}_{b_{i}}$ stands for the
covariant derivative $\ $and $\widetilde{\operatorname{div}}$ stands for the
covariant divergence, see formulas (\ref{covder}) and
(\ref{divergenceRiemannian}) in Appendix \ref{SubSectDG}. Notice that we are
using Leibnitz' rule for covariant derivatives, and the fact that covariant
derivatives of a scalar field are just partial derivatives.

Next, we apply the Divergence Theorem for $\Gamma$ as formulated in Appendix
\ref{SubSectDG}, formula (\ref{StokesRiemannian}). Since $\Gamma$ has empty
boundary we readily obtain%
\begin{equation}
\int_{\Gamma}\widetilde{\operatorname{div}}G_{\mathrm{B}}\,\mathrm{dS}%
_{\mathrm{b}}=0, \label{StokesGamma}%
\end{equation}
and can finally rewrite (\ref{ActionMulti}) as%
\begin{gather*}
\delta S=\int_{t_{0}}^{t_{1}}\int_{\Gamma}%
{\displaystyle\sum\nolimits_{\ell}}
\left[  \left\langle G_{\mathrm{D}}^{\ell}(x(b),t)\,,\mathrm{n}%
(b)\right\rangle \,+\partial_{\psi^{\ell}}L_{\mathrm{INT}}(t,(\psi
_{\mathrm{L}}^{\ell}-\psi_{\mathrm{B}}^{\ell})(b,t))\right]  \delta
\psi_{\mathrm{L}}^{\ell}\,\mathrm{dS}_{\mathrm{b}}\mathrm{dt+}\\
+\int_{t_{0}}^{t_{1}}\int_{\Gamma}%
{\displaystyle\sum\nolimits_{\ell}}
\left[  \frac{\partial L_{\mathrm{B}}}{\partial\psi_{\mathrm{B}}^{\ell}%
}-\partial_{0}G_{\mathrm{B}}^{\ell0}\,-\widetilde{\operatorname{div}%
}G_{\mathrm{B,}j}\right]  \,\delta\psi_{\mathrm{B}}^{\ell}\,\mathrm{dS}%
_{\mathrm{b}}\mathrm{dt-}\\
-\int_{t_{0}}^{t_{1}}\int_{\Gamma}\left[  \partial_{\psi}L_{\mathrm{INT}%
}(t,(\psi_{\mathrm{L}}^{\ell}-\psi_{\mathrm{B}}^{\ell})(b,t))\right]
\cdot\delta\psi_{\mathrm{B}}^{\ell}\,\mathrm{dS}_{\mathrm{b}}\mathrm{dt},
\end{gather*}
where we have taken into account that $\psi_{\mathrm{L}}^{\ell}(b,t)=\psi
_{\mathrm{D}}^{\ell}(x(b),t)$. Arbitrariness and independence of $\delta
\psi_{\mathrm{L}}^{\ell}$ and $\delta\psi_{\mathrm{B}}^{\ell}$ then entail the
BEL equations (\ref{MBEL}):%
\begin{gather}
\text{(IEL)\qquad}\qquad\left\langle G_{\mathrm{D}}^{\ell}%
(x(b),t),\,\mathrm{n}(b)\right\rangle +\partial_{\psi^{\ell}}L_{\mathrm{INT}%
}=0;\qquad b\in\Gamma,t>0;\label{BELMulti}\\
\text{(BEL)\qquad}\frac{\partial L_{\mathrm{B}}}{\partial\psi_{\mathrm{B}%
}^{\ell}}-\widetilde{\operatorname{div}}G_{\mathrm{B}}^{\ell}-\partial
_{0}G_{\mathrm{B}}^{\ell0}-\partial_{\psi^{\ell}}L_{\mathrm{INT}}=0;\qquad
b\in\Gamma,t>0\nonumber
\end{gather}
for each $\ell=1,2,...k$.

\subsubsection{Lower dimensional boundaries\label{LowerDimBound}}

\begin{figure}
[h]
\begin{center}
\ifcase\msipdfoutput
\includegraphics[
trim=0.000000in 1.407827in 0.000000in 0.170493in,
height=3.2146in,
width=6.1478in
]%
{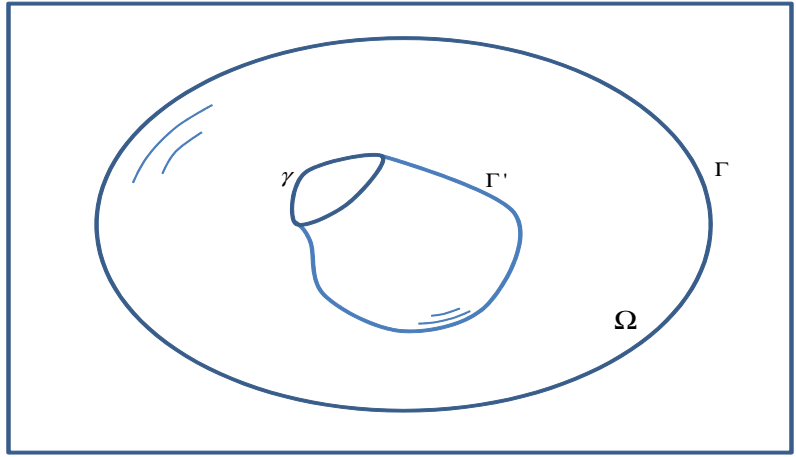}%
\else
\includegraphics[
height=3.2146in,
width=6.1478in
]%
{D:/Initial-BoundaryValueProblem/ArXiv/graphics/MultipleBoundaries__3.pdf}%
\fi
\caption{$\Gamma^{\prime}$ is an "interior", non-closed surface, with
one-dimensional boundary $\gamma.$}%
\label{MultipleBoundaries}%
\end{center}
\end{figure}

Apart from the usual $(n-1)$-dimensional \ closed hypersurface, $n$%
-dimensional systems can exhibit boundary subsystems, defined on lower
dimensional submanifolds. Suppose for example that our main system is defined
on a three dimensional region $\Omega,$ and, apart from the closed "external"
boundary $\Gamma,$ there is an interior piece of the boundary $\Gamma^{\prime
}$ which is not a closed surface but one with non-empty, closed,
one-dimensional boundary $\gamma:$ $r\in\lbrack0,1]\rightarrow E^{3}$, see
Figure \ref{MultipleBoundaries}.%

Then, independent boundary fields $\psi_{\mathrm{B}}^{\ell\prime}$ can be
defined on $\gamma,$ governed by a one-dimensional Lagrangian density
\[
L_{\mathrm{B}}^{\prime}(t,\psi_{\mathrm{B}}^{\ell\prime},D\psi_{\mathrm{B}%
}^{\ell\prime}),
\]
as well as a new interaction Lagrangian which, according to the assumption
stated in Subsection \ref{SubsectBoundInt}, has the form%
\begin{equation}
L_{\mathrm{INT}}^{\prime}(t,\psi_{\mathrm{B}}^{\ell\prime}-\psi_{\mathrm{L}%
}^{\ell\prime}), \label{IntLagrangian2}%
\end{equation}
where
\[
\psi_{\mathrm{L}}^{\ell\prime}(r,t)=\lim_{b\rightarrow b(r)}\psi_{\mathrm{B}%
}^{\ell}(b,t).
\]
Then, when we deal with the "bulk" system, defined on $\Omega,$ the term
containing the divergence may in principle contribute two different surface
integrals on $\Gamma^{\prime},$ since $\psi_{\mathrm{D}}^{\ell}$ might be
discontinuous across $\Gamma^{\prime}$. These surface integrals add to the
non-divergent part of the integral arising when we deal with the boundary
subsystem defined on $\Gamma^{\prime}.$ We get, apart from the equation that
holds in the volume, two equations corresponding to each of the pieces of the
boundary, $\Gamma$ and $\Gamma^{\prime},$ since we can vary independently
$\psi_{\mathrm{L}}^{\ell}$ and $\psi_{\mathrm{B}}$ on both $\Gamma$ and
$\Gamma^{\prime}$. The integral of the divergence on $\Gamma^{\prime}$ will
not vanish, but equal%
\begin{equation}
\int_{\gamma}%
{\displaystyle\sum\nolimits_{\ell}}
\left\langle G_{\mathrm{B}}^{\ell}(b(r),t)\,,\widetilde{\mathrm{n}%
}(b(r))\right\rangle \,\delta\psi_{\mathrm{B}}^{\ell}(b(r),t)\,\mathrm{dl}%
_{\mathrm{r}}=\int_{\gamma}%
{\displaystyle\sum\nolimits_{\ell}}
\left\langle G_{\mathrm{B}}^{\ell}(b(r),t)\,,\widetilde{\mathrm{n}%
}(b(r))\right\rangle \,\delta\psi_{\mathrm{L}}^{\ell^{\prime}}\,\mathrm{dl}%
_{\mathrm{r}}\mathrm{,} \label{newterm}%
\end{equation}
where $\mathrm{dl}_{\mathrm{r}}=\left\vert \frac{dl}{dr}\right\vert
\mathrm{dr}$ is the length element on $\gamma$ and $\widetilde{\mathrm{n}}(r)$
is the unit normal vector to $\gamma$ in the tangent space to $\Gamma^{\prime
}$(such vector is uniquely determined by the Riemannian structure of
$\Gamma^{\prime}$), see formula (\ref{StokesGamma}) in Appendix
\ref{SubSectDG}. The term (\ref{newterm}) interacts with the non-vanishing
term arising from the variation of the action associated to $L_{\mathrm{B}%
}^{\prime},$ giving rise to two one-dimensional IEL-BEL equations, as
$\psi_{\mathrm{B}}$ and $\psi_{\mathrm{B}}^{\prime}$ are varied independently.
The latter equations also contain terms of the form%
\[
\partial_{\psi^{\prime}}L_{\mathrm{INT}}^{\prime}(t,(\psi_{\mathrm{L}}%
^{\prime}-\psi_{\mathrm{B}}^{\prime})(r,t)),
\]
arising from the new interaction Lagrangian (\ref{IntLagrangian2}). We leave
the details to the interested reader.

The above procedure can be applied without significant changes to fields
defined on subsets of $\mathbb{E}^{n}$ with any number (as allowed by
dimension) of lower-dimensional boundary fields. It should be noted that the
interaction between the field on the $k$-dimensional boundary and the one on
the $(k-1)$-dimensional boundary takes place via the \textsl{normal
}derivatives of the former. The geometry of the boundaries is reflected
through terms containing $g^{(s)}=\det g_{\alpha\beta}^{(s)}$, $s=k-1,k$,
where $g_{\alpha\beta}^{(s)}$ stands for the metrics induced on the
$s$-dimensional manifold.

\section{Energy distribution and flow\label{SectEnergy}}

The Lagrangian set up of our system with two Lagrangian densities
$L_{\mathrm{D}}$ and $L_{\mathrm{B}}+L_{\mathrm{INT}}$ allows to obtain a
detailed picture of the energy distribution inside system domain and its
boundary\ as well as the energy transport between them. The quantitative
treatment of the corresponding energy densities $H_{\mathrm{D}}$ and
$H_{\mathrm{B}}$ and energy fluxes $S_{\mathrm{D}}$ and $S_{\mathrm{B}}$ can
be handled using the method described in Section \ref{apencon} to construct
and justify conventional expressions for the energy densities $H_{\mathrm{D}}$
and $H_{\mathrm{B}}$ and fluxes $S_{\mathrm{D}}$ and $S_{\mathrm{B}}$ based on
the corresponding Lagrangian densities $L_{\mathrm{D}}$ and $L_{\mathrm{B}%
}+L_{\mathrm{INT}}$. That might seem to be obvious but the presence of the
limit value $\psi_{\mathrm{L}}^{\ell}$ of $\psi_{\mathrm{D}}^{\ell}$ in the
interaction Lagrangian $L_{\mathrm{INT}}$ raises a concern on the legitimacy
of the usage of conventional expressions derived for conventional Lagrangian
densities. At conceptual level one can argue that the presence of the limit
value $\psi_{\mathrm{L}}^{\ell}$ in the boundary Lagrangian is similar to an
ideal holonomic constraint associated always with reaction forces that do no
work, \cite[III.1]{LanVPM}, \cite[1.2]{Gantmacher}. Consequently, one can
imply that though the presence of a holonomic constraint affects solutions to
the Euler-Lagrange equations it should have no effect on expressions for the
energy densities and energy fluxes. That is true indeed as we verify below by
a direct computation of the energy densities, energy fluxes and detailed
energy conservation laws at every point in the domain interiors or in its boundary.

\subsection{One-dimensional domain\label{SubsectEnergyOneDim}}

In one-dimensional case the system is described by Lagrangian densities
$L_{\mathrm{D}}$ and $L_{\mathrm{B}}+L_{\mathrm{INT,1}}+L_{\mathrm{INT,2}}$.
Let us start with the interval Lagrangian $L_{\mathrm{D}}$ and introduce the
following conventional expressions for the energy density and energy flux
\begin{align}
H_{\mathrm{D}}  &  =%
{\displaystyle\sum\nolimits_{\ell}}
\frac{\partial L_{\mathrm{D}}}{\partial\partial_{t}\psi_{\mathrm{D}}^{\ell}%
}\partial_{t}\psi_{\mathrm{D}}^{\ell}-L_{\mathrm{D}}\text{ is the energy
density,}\label{oneHSD1}\\
S_{\mathrm{D}}  &  =%
{\displaystyle\sum\nolimits_{\ell}}
\frac{\partial L_{\mathrm{D}}}{\partial\partial_{z}\psi_{\mathrm{D}}^{\ell}%
}\partial_{t}\psi_{\mathrm{D}}^{\ell},\quad\text{ is the energy flux.}
\label{oneHSD2}%
\end{align}
The conventional expressions for the energy at boundary points $b_{i}$,
$i=1,2$ are%
\begin{equation}
H_{\mathrm{B}}\left(  b_{i},t\right)  =%
{\displaystyle\sum\nolimits_{\ell}}
\frac{\partial L_{\mathrm{B}}}{\partial\partial_{t}\psi_{\mathrm{B}}^{\ell
}\left(  b_{i}\right)  }\partial_{t}\psi_{\mathrm{B}}^{\ell}\left(
b_{i}\right)  -\left(  L_{\mathrm{B}}+L_{\mathrm{INT,i}}\right)  \text{.}
\label{oneHSD3}%
\end{equation}
Assume that $\left\{  \psi_{\mathrm{D}}^{\ell},\psi_{\mathrm{B}}^{\ell}%
,\psi_{\mathrm{L}}^{\ell}\right\}  $ is a solution to the EL equations
(\ref{DEL})-(\ref{BEL2}). Then (\ref{DEL}) readily implies the following
differential form of the energy conservation law in the interval $(b_{1}%
,b_{2}):$
\begin{equation}
\frac{dH_{\mathrm{D}}}{dt}+\partial_{z}S_{\mathrm{D}}=-\partial_{t}%
L_{\mathrm{D}}. \label{oneHSD4}%
\end{equation}
The integral form of the above conservation is%
\begin{equation}
\frac{d}{dt}\int_{b_{1}}^{b_{2}}H_{\mathrm{D}}\,\mathrm{dx}+%
{\displaystyle\sum\nolimits_{\ell}}
\left(  G_{\mathrm{D}}^{\ell}\partial_{t}\psi_{\mathrm{L}}^{\ell}\left(
b_{2}\right)  -G_{\mathrm{D}}^{\ell}\partial_{t}\psi_{\mathrm{L}}^{\ell
}\left(  b_{1}\right)  \right)  =-\int_{b_{1}}^{b_{2}}\partial_{t}%
L_{\mathrm{D}}\,\mathrm{dx.} \label{oneHSD5}%
\end{equation}
Let us turn now to the boundary points $b_{1}$ and $b_{2}$ and their
respective Lagrangians $L_{\mathrm{B}}+L_{\mathrm{INT,1}}$ and $L_{\mathrm{B}%
}+L_{\mathrm{INT,2}}$ where $\psi_{\mathrm{L}}^{\ell}\left(  b_{1},t\right)  $
and $\psi_{\mathrm{L}}^{\ell}\left(  b_{2},t\right)  $ are components of a
solution $\left\{  \psi_{\mathrm{D}}^{\ell},\psi_{\mathrm{B}}^{\ell}%
,\psi_{\mathrm{L}}^{\ell}\right\}  $ to the complete set of the EL equations.
Let us assume now that those $\psi_{\mathrm{L}}^{\ell}\left(  b_{1},t\right)
$ and $\psi_{\mathrm{L}}^{\ell}\left(  b_{2},t\right)  $ are fixed. Notice
then that $\left\{  \psi_{\mathrm{B}}^{\ell}\left(  b_{1},t\right)  \right\}
$ and $\left\{  \psi_{\mathrm{B}}^{\ell}\left(  b_{2},t\right)  \right\}  $
are solutions to the EL equations (\ref{BEL1}) and (\ref{BEL2}) respectively
associated with the corresponding Lagrangian $L_{\mathrm{B}}+L_{\mathrm{INT,1}%
}$ and $L_{\mathrm{B}}+L_{\mathrm{INT,2}}$ with fixed $\psi_{\mathrm{L}}%
^{\ell}$. Based on this we readily obtain the following energy conservation
law at the boundary points%
\begin{equation}
\frac{dH_{\mathrm{B}}\left(  b_{i},t\right)  }{dt}=-\partial_{t}L_{\mathrm{B}%
}-%
{\displaystyle\sum\nolimits_{\ell}}
\partial_{\psi^{\ell}}\left(  L_{\mathrm{INT,i}}\right)  \partial_{t}%
\psi_{\mathrm{L}}^{\ell}\left(  b_{i},t\right)  ,\quad i=1,2. \label{oneHSD6}%
\end{equation}
Multiplying the IEL equations (\ref{IEL}) by $\psi_{\mathrm{L}}^{\ell}\left(
b_{1},t\right)  \ $(respectively $\psi_{\mathrm{L}}^{\ell}\left(
b_{2},t\right)  $) we obtain%

\begin{align}
-G_{\mathrm{D}}^{\ell}\partial_{t}\psi_{\mathrm{L}}^{\ell}\left(
b_{1},t\right)  +\partial_{\psi^{\ell}}\left(  L_{\mathrm{INT,1}}\right)
\partial_{t}\psi_{\mathrm{L}}^{\ell}\left(  b_{1},t\right)   &
=0,\label{oneHSD7}\\
G_{\mathrm{D}}^{\ell}\partial_{t}\psi_{\mathrm{L}}^{\ell}\left(
b_{2},t\right)  +\partial_{\psi^{\ell}}\left(  L_{\mathrm{INT,1}}\right)
\partial_{t}\psi_{\mathrm{L}}^{\ell}\left(  b_{2},t\right)   &  =0.\nonumber
\end{align}
The above equations signify an exact balance between the energy flux
$S_{\mathrm{D}}^{\ell}\left(  b,t\right)  =G_{\mathrm{D}}^{\ell}\partial
_{t}\psi_{\mathrm{D}}^{\ell}\left(  b,t\right)  =G_{\mathrm{D}}^{\ell}%
\partial_{t}\psi_{\mathrm{L}}^{\ell}\left(  b,t\right)  $ associated with the
domain and the similar quantity $\partial_{\psi^{\ell}}\left(  L_{\mathrm{INT}%
}\right)  \partial_{t}\psi_{\mathrm{L}}^{\ell}$ associated with the boundary.
Combining the equations (\ref{oneHSD5})-(\ref{oneHSD7}) we obtain the
following total energy conservation law%
\begin{gather}
\frac{d}{dt}\left[  \int_{b_{1}}^{b_{2}}H_{\mathrm{D}}\,\mathrm{dx}%
+H_{\mathrm{B}}\left(  b_{1}\right)  +H_{\mathrm{B}}\left(  b_{2}\right)
\right]  =-\int_{b_{1}}^{b_{2}}\partial_{t}L_{\mathrm{D}}\,\mathrm{dx}%
-\label{oneHSD8}\\
-\partial_{t}\left[  \left(  L_{\mathrm{B}}+L_{\mathrm{INT,1}}\right)  \left(
b_{1}\right)  +\left(  L_{\mathrm{B}}+L_{\mathrm{INT,2}}\right)  \left(
b_{2}\right)  \right]  .\nonumber
\end{gather}
If external forces with density $F_{\mathrm{D}}^{\ell}$ distributed on the
interval $(b_{1},b_{2})$ or forces $F_{\mathrm{B}}^{\ell}$ on the boundary are
present, then the individual (bulk and boundary) conservation laws are
suitably modified, see Section \ref{apencon}. In this case, (\ref{oneHSD4})
becomes%
\[
\frac{dH_{\mathrm{D}}}{dt}+\partial_{z}S_{\mathrm{D}}=-\partial_{t}%
L_{\mathrm{D}}+%
{\displaystyle\sum\nolimits_{\ell}}
F_{\mathrm{D}}^{\ell}\partial_{t}\psi_{\mathrm{D}}^{\ell}%
\]
with the corresponding modification of (\ref{oneHSD5}). In turn,
(\ref{oneHSD6}) takes the form%
\begin{equation}
\frac{dH_{\mathrm{B}}\left(  b_{i},t\right)  }{dt}=-\partial_{t}L_{\mathrm{B}%
}-%
{\displaystyle\sum\nolimits_{\ell}}
\partial_{\psi^{\ell}}\left(  L_{\mathrm{INT,i}}\right)  \partial_{t}%
\psi_{\mathrm{L}}^{\ell}\left(  b_{i},t\right)  +%
{\displaystyle\sum\nolimits_{\ell}}
F_{\mathrm{B}}^{\ell}\partial_{t}\psi_{\mathrm{B}}^{\ell}\left(
b_{i},t\right)  ,\qquad i=1,2.
\end{equation}
Equations (\ref{oneHSD7}) are not modified by external forces. Finally, the
combined integral form (\ref{oneHSD8}) takes the form%
\begin{gather}
\frac{d}{dt}\left[  \int_{b_{1}}^{b_{2}}H_{\mathrm{D}}\,\mathrm{dx}%
+H_{\mathrm{B}}\left(  b_{1}\right)  +H_{\mathrm{B}}\left(  b_{2}\right)
\right]  =-\int_{b_{1}}^{b_{2}}\partial_{t}L_{\mathrm{D}}\,\mathrm{dx}\\
-\partial_{t}\left[  \left(  L_{\mathrm{B}}+L_{\mathrm{INT,1}}\right)  \left(
b_{1}\right)  +\left(  L_{\mathrm{B}}+L_{\mathrm{INT,2}}\right)  \left(
b_{2}\right)  \right]  +%
{\displaystyle\sum\nolimits_{\ell}}
\left[  \int_{b_{1}}^{b_{2}}F_{\mathrm{D}}^{\ell}\partial_{t}\psi_{\mathrm{D}%
}^{\ell}\mathrm{dx}+F_{\mathrm{B}}^{\ell}\partial_{t}\psi_{\mathrm{B}}^{\ell
}\right]  \mathrm{,}\nonumber
\end{gather}
containing the total power dissipated by external forces.

\subsection{Multidimensional domain\label{SubsectEnergyMultidim}}

In the multidimensional case the system is described by Lagrangian densities
$L_{\mathrm{D}}$ and $L_{\mathrm{B}}+L_{\mathrm{INT}}$ associated respectively
with the domain and its boundary. Let us start with the domain Lagrangian
$L_{\mathrm{D}}.$ Using the notations introduced in (\ref{Gs}), we can
introduce the following conventional expressions for the energy density and
the energy flux
\begin{align}
H_{\mathrm{D}}  &  =%
{\displaystyle\sum\nolimits_{\ell}}
G_{\mathrm{D}}^{\ell0}\partial_{t}\psi_{\mathrm{D}}^{\ell}-L_{\mathrm{D}%
}\text{ is the energy density,}\label{muGDpsi3}\\
S_{\mathrm{D}}^{i}  &  =%
{\displaystyle\sum\nolimits_{\ell}}
G_{\mathrm{D}}^{\ell i}\partial_{t}\psi_{\mathrm{D}}^{\ell},\quad i=1,\ldots
n,\text{ is the energy flux vector.} \label{muGDpsi4}%
\end{align}
We also use the following concise presentation for the energy flux vector
\begin{equation}
S_{\mathrm{D}}=%
{\displaystyle\sum\nolimits_{\ell}}
S_{\mathrm{D}}^{\ell},\quad S_{\mathrm{D}}^{\ell}=G_{\mathrm{D}}^{\ell
}\partial_{t}\psi_{\mathrm{D}}^{\ell}. \label{muGDpsi5}%
\end{equation}
The conventional expressions for the energy density and the energy flux on the
boundary are%
\begin{align}
H_{\mathrm{B}}  &  =%
{\displaystyle\sum\nolimits_{\ell}}
G_{\mathrm{D}}^{\ell0}\partial_{t}\psi_{\mathrm{B}}^{\ell}-\left(
L_{\mathrm{B}}+L_{\mathrm{INT}}\right)  \text{ is the energy density,}%
\label{muGDpsi8}\\
S_{\mathrm{B}}^{i}  &  =%
{\displaystyle\sum\nolimits_{\ell}}
G_{\mathrm{B}}^{\ell i}\partial_{t}\psi_{\mathrm{B}}^{\ell},\quad i=1,\ldots
n\text{ is the energy flux,} \label{muGDpsi9}%
\end{align}%
\begin{equation}
S_{\mathrm{B}}=%
{\displaystyle\sum\nolimits_{\ell}}
S_{\mathrm{B}}^{\ell},\quad S_{\mathrm{B}}^{\ell}=G_{\mathrm{B}}^{\ell
}\partial_{t}\psi_{\mathrm{B}}^{\ell}. \label{muGDpsi10}%
\end{equation}

Assume that $\left\{  \psi_{\mathrm{D}}^{\ell},\psi_{\mathrm{B}}^{\ell}%
,\psi_{\mathrm{L}}^{\ell}\right\}  $ is a solution to the EL equations
(\ref{MDEL}), (\ref{MIEL}), (\ref{MBEL}). Then evidently $\left\{
\psi_{\mathrm{D}}^{\ell}\right\}  $ is a solution to the EL equation
(\ref{MDEL}) associated with the Lagrangian $L_{\mathrm{D}}$ and, according to
the argument of Section \ref{apencon}, the following energy conservation law
holds inside the domain $\Omega$
\begin{equation}
\frac{dH_{\mathrm{D}}}{dt}+\operatorname{div}S_{\mathrm{D}}=-\partial
_{t}L_{\mathrm{D}}. \label{HDSDt1}%
\end{equation}
The above differential form of the energy conservation law combined with the
definition (\ref{muGDpsi4}), (\ref{muGDpsi5}) of the energy flux
$S_{\mathrm{D}}^{\ell}=G_{\mathrm{D}}^{\ell}\partial_{t}\psi_{\mathrm{D}%
}^{\ell}$ imply the following integral form of the energy conservation%
\begin{equation}
\frac{d}{dt}\int_{\Omega}H_{\mathrm{D}}\,\mathrm{dx}+%
{\displaystyle\sum\nolimits_{\ell}}
\int_{\Gamma}\left\langle G_{\mathrm{D}}^{\ell},\mathrm{n}\left(  b\right)
\right\rangle \partial_{t}\psi_{\mathrm{L}}^{\ell}\,\mathrm{dS}_{\mathrm{b}%
}=-\int_{\Omega}\partial_{t}L_{\mathrm{D}}\,\mathrm{dx,} \label{HDSDt2}%
\end{equation}
where $\mathrm{n}\left(  b\right)  $ is the normal vector to boundary $\Gamma$
at a point $b$. Notice that we used also the definition of $\psi_{\mathrm{L}%
}^{\ell}$ as the limit value of $\psi_{\mathrm{D}}^{\ell}$, that is
\begin{equation}
\psi_{\mathrm{D}}^{\ell}\left(  b\right)  =\psi_{\mathrm{L}}^{\ell}\left(
b\right)  ,\quad b\in\Gamma. \label{HDSDt3}%
\end{equation}

In the case of the boundary let us consider the Lagrangian $L_{\mathrm{B}%
}+L_{\mathrm{INT}}$ where $\psi_{\mathrm{L}}^{\ell}$ is a component of a
solution $\left\{  \psi_{\mathrm{D}}^{\ell},\psi_{\mathrm{B}}^{\ell}%
,\psi_{\mathrm{L}}^{\ell}\right\}  $ to the EL equations (\ref{MDEL}),
(\ref{MIEL}), (\ref{MBEL}). Suppose now that those $\psi_{\mathrm{L}}^{\ell}$
are fixed. Notice then that $\left\{  \psi_{\mathrm{B}}^{\ell}\right\}  $ is a
solution to the EL equation (\ref{MBEL}) associated with the Lagrangian
$L_{\mathrm{B}}+L_{\mathrm{INT}}$ with $\psi_{\mathrm{L}}^{\ell}$ being fixed.
Using once again the argument of Section \ref{apencon} we obtain the following
conservation law
\begin{equation}
\frac{dH_{\mathrm{B}}}{dt}+\widetilde{\operatorname{div}}S_{\mathrm{B}%
}=-\partial_{t}\left(  L_{\mathrm{B}}+L_{\mathrm{INT}}\right)  -%
{\displaystyle\sum\nolimits_{\ell}}
\partial_{\psi^{\ell}}\left(  L_{\mathrm{INT}}\right)  \partial_{t}%
\psi_{\mathrm{L}}^{\ell}, \label{HDSDt4}%
\end{equation}
Integrating the above equation over the boundary of the Riemann manifold
$\Gamma$ which has no boundary and using the Riemannian version of Gauss
Theorem, (\ref{StokesRiemannian}) we obtain the integral form of the energy
conservation on the boundary%
\begin{equation}
\frac{d}{dt}\int_{\Gamma}H_{\mathrm{B}}\,\mathrm{dS}_{\mathrm{b}}+\int
_{\Gamma}%
{\displaystyle\sum\nolimits_{\ell}}
\partial_{\psi^{\ell}}\left(  L_{\mathrm{INT}}\right)  \partial_{t}%
\psi_{\mathrm{L}}^{\ell}\,\mathrm{dS}_{\mathrm{b}}=-\int_{\Gamma}\partial
_{t}\left(  L_{\mathrm{B}}+L_{\mathrm{INT}}\right)  \,\mathrm{dS}_{\mathrm{b}%
}\mathrm{,} \label{HDSDt5}%
\end{equation}
Observe now that by multiplying the interface IEL equation (\ref{MIEL}) by
$\partial_{t}\psi_{\mathrm{L}}^{\ell}$ we obtain
\begin{equation}
\left\langle G_{\mathrm{D}}^{\ell},\mathrm{n}\left(  b\right)  \right\rangle
\partial_{t}\psi_{\mathrm{L}}^{\ell}+\partial_{\psi^{\ell}}\left(
L_{\mathrm{INT}}\right)  \partial_{t}\psi_{\mathrm{L}}^{\ell}=0.
\label{HDSDt6}%
\end{equation}
The above equation signifies an exact balance between the normal component of
energy flux $S_{\mathrm{D}}^{\ell}\left(  b\right)  =G_{\mathrm{D}}^{\ell
}\partial_{t}\psi_{\mathrm{D}}^{\ell}\left(  b\right)  =G_{\mathrm{D}}^{\ell
}\partial_{t}\psi_{\mathrm{L}}^{\ell}\left(  b\right)  $ associated with the
domain and the similar quantity $\partial_{\psi^{\ell}}\left(  L_{\mathrm{INT}%
}\right)  \partial_{t}\psi_{\mathrm{L}}^{\ell}$ associated with the boundary.
We refer to equality (\ref{HDSDt6}) as the \emph{detailed energy balance} for
it is satisfied at every point point $b$ of the boundary $\Gamma$.

Combining now equalities (\ref{HDSDt4})-(\ref{HDSDt6}) we obtain the total
energy conservation%
\begin{equation}
\frac{d}{dt}\left[  \int_{\Omega}H_{\mathrm{D}}\,\mathrm{dx}+\int_{\Gamma
}H_{\mathrm{B}}\,\mathrm{dS}_{\mathrm{b}}\right]  =-\int_{\Omega}\partial
_{t}L_{\mathrm{D}}\,\mathrm{dx}-\int_{\Gamma}\partial_{t}\left(
L_{\mathrm{B}}+L_{\mathrm{INT}}\right)  \mathrm{dS}_{\mathrm{b}}\mathrm{,}
\label{HDSDt7}%
\end{equation}
signifying the balance of the time derivative of the total energy residing in
the domain and its boundary and the power generated by external sources.

If external forces on the volume and/or on the boundary are present, with
densities $F_{\mathrm{D}}^{\ell}$ and $F_{\mathrm{B}}^{\ell},$ equation
\ref{HDSDt1} becomes%
\[
\frac{dH_{\mathrm{D}}}{dt}+\operatorname{div}S_{\mathrm{D}}=-\partial
_{t}L_{\mathrm{D}}+%
{\displaystyle\sum\nolimits_{\ell}}
F_{\mathrm{D}}^{\ell}\partial_{t}\psi_{\mathrm{D}}^{\ell}%
\]
while \ref{HDSDt4} becomes%
\begin{equation}
\frac{dH_{\mathrm{B}}}{dt}+\widetilde{\operatorname{div}}S_{\mathrm{B}%
}=-\partial_{t}\left(  L_{\mathrm{B}}+L_{\mathrm{INT}}\right)  -%
{\displaystyle\sum\nolimits_{\ell}}
\partial_{\psi^{\ell}}\left(  L_{\mathrm{INT}}\right)  \partial_{t}%
\psi_{\mathrm{L}}^{\ell}+%
{\displaystyle\sum\nolimits_{\ell}}
F_{\mathrm{B}}^{\ell}\partial_{t}\psi_{\mathrm{B}}^{\ell}. \label{HDSDt8}%
\end{equation}
Finally, the combined integral form \ref{HDSDt7} takes the form%
\begin{gather}
\frac{d}{dt}\left[  \int_{\Omega}H_{\mathrm{D}}\,\mathrm{dx}+\int_{\Gamma
}H_{\mathrm{B}}\,\mathrm{dS}_{\mathrm{b}}\right]  =\label{HDSDt9}\\
=-\int_{\Omega}\partial_{t}L_{\mathrm{D}}\,\mathrm{dx}-\int_{\Gamma}%
\partial_{t}\left(  L_{\mathrm{B}}+L_{\mathrm{INT}}\right)  \mathrm{dS}%
_{\mathrm{b}}+%
{\displaystyle\sum\nolimits_{\ell}}
\left[  \int_{\Omega}F_{\mathrm{D}}^{\ell}\partial_{t}\psi_{\mathrm{D}}^{\ell
}\mathrm{dx+}\int_{\Gamma}F_{\mathrm{B}}^{\ell}\partial_{t}\psi_{\mathrm{B}%
}^{\ell}\mathrm{dS}_{\mathrm{b}}\right]  \mathrm{,}\nonumber
\end{gather}
containing the total power dissipated by external forces.

\section{Some examples\label{SectExamples}}

In this section we present applications of our approach. In Subsection
\ref{SubSectOneDimExamples} we describe a sequence of increasingly complex
conservative one-dimensional mechanical systems, composed by masses, springs
and a unique string, which in all cases represents the "interior" continuous
system. In Subsection \ref{SubSectionMTL}, we present an application to
multi-transmission lines, in which the carrier of the field is still
one-dimensional but the field itself is multidimensional. Subsection
\ref{SubSectMembrane} deals with a two-dimensional application.

\subsection{One dimensional examples\label{SubSectOneDimExamples}}

This subsection deals with one-dimensional examples in which the "bulk" system
is a semi-infinite string. It should be noticed that often we look at the
system as an observer residing on the boundary rather than as one residing in
the system interior. This point of view differs from the conventional one and
arises from the treatment of the interior and the boundary systems on an equal footing.

\subsubsection{String attached to a spring\label{SubSubSimpleSpring}}

Our simplest example is that of a semi-infinite string attached to a
(massless) spring at its end, see Figure \ref{Figure1} (a). Suppose the string
is stretched along the $z-$interval $[0,\infty)$ and small transversal
oscillations propagate along it (hence we are within the linear model). The
spring is assumed linear and is attached to the end of the string at $z=0$ in
such a way, that only transversal oscillations are allowed. We consider the
string as the "bulk" system and $\psi_{\mathrm{D}}=\psi_{\mathrm{D}}(z,t)$
stands its deflection, whereas the attached spring is the boundary system and
$\psi_{\mathrm{B}}(t)$ is its vertical displacement from the equilibrium. It
is clear that $\psi_{\mathrm{B}}(t)=$ $\psi_{\mathrm{L}}(t)=\psi_{\mathrm{D}%
}(0,t)$ (continuity constraint) and the bulk, boundary and interaction
Lagrangian densities are, respectively,%
\[
L_{\mathrm{D}}=\frac{\rho}{2}(\mathcal{\partial}_{t}\psi_{\mathrm{D}}%
)^{2}-\frac{T}{2}(\mathcal{\partial}_{z}\psi_{\mathrm{D}})^{2};\quad
L_{\mathrm{B}}=-\frac{k}{2}\psi_{\mathrm{B}}^{2};\quad L_{\mathrm{INT}}=0,
\]
where $k>0$ is the Hooke constant of the spring, $T$ is the constant tension
and $\rho$ is the linear density of the string. The total Lagrangian of the
system is%
\[
L\mathcal{(\partial}_{t}\psi_{\mathrm{D}},\mathcal{\partial}_{z}%
\psi_{\mathrm{D}},\psi_{\mathrm{B}})=\frac{1}{2}\int_{0}^{\infty}\left[
\rho(\mathcal{\partial}_{t}\psi_{\mathrm{D}})^{2}-T(\mathcal{\partial}_{z}%
\psi_{\mathrm{D}})^{2}\right]  \,\mathrm{dz}-\frac{k}{2}\psi_{\mathrm{B}}%
^{2},
\]
This case is well known in literature, see e.g. \cite{FG} and we omit the
details. An application of our general formulas (\ref{DEL})-(\ref{BEL2})
produces the wave equation%
\begin{equation}
\mathcal{\partial}_{tt}^{2}\psi_{\mathrm{D}}=a^{2}\mathcal{\partial}_{zz}%
^{2}\psi_{\mathrm{D}};\qquad a^{2}=\frac{T}{\rho} \label{waveeq}%
\end{equation}
for the evolution of the string, and the well-known Robin boundary condition%
\begin{equation}
T\mathcal{\partial}_{z}\psi_{\mathrm{D}}(0,t)-k\psi_{\mathrm{B}}(t)=0,\qquad
t\geq0 \label{BCsimple}%
\end{equation}
which follows from the first equation in (\ref{GenOneDimBC}) and the
continuity constraint. Observe that in this example one must first include an
interaction term and take the limit as its strength tends to infinity,
according to the discussion at the end of subsection \ref{MainResult1D}. \ See
the example considered in subsection \ref{SubSubLambModified}, where we carry
out a similar limit process explicitly.

If we assume that the string has been at rest for all $t\leq0,$ in the general
solution of (\ref{waveeq}), given by d'Alembert's formula%
\[
\psi_{\mathrm{D}}(z,t)=f(z-at)+g(z+at),
\]
necessarily $g\equiv0$ (no backward wave) and then $\psi_{\mathrm{D}}$
satisfies the one-directional wave equation%
\[
\mathcal{\partial}_{t}\psi_{\mathrm{D}}+a\mathcal{\partial}_{z}\psi
_{\mathrm{D}}=0.
\]
We can then replace $\mathcal{\partial}_{z}\psi_{\mathrm{D}}%
(0,t)=-\mathcal{\partial}_{t}\psi_{\mathrm{D}}(0,t)/a=-\mathcal{\partial}%
_{t}\psi_{\mathrm{B}}(t)/a$ in (\ref{BCsimple}). This leads to the boundary
evolution equation%
\begin{equation}
\mathcal{\partial}_{t}\psi_{\mathrm{B}}+\frac{ak}{T}\psi_{\mathrm{B}}=0,\qquad
t\geq0. \label{LambNoMass}%
\end{equation}
If an initial impulsive force is applied at $z=0$, a pulse is propagated along
the string, taking the energy to infinity. The boundary system returns
exponentially fast to equilibrium.

Observe that imposing the absence of a backward wave is a kind of "boundary
condition" at infinity, usually called "non-radiation" condition. If we assume
non-zero initial conditions for the system%
\[
\psi_{\mathrm{D}}(z,0)=g(z);\qquad\partial_{t}\psi_{\mathrm{D}}(z,0)=h(z),
\]
it can be easily proved that they enter the boundary evolution equation as
sources%
\[
\mathcal{\partial}_{t}\psi_{\mathrm{B}}+\frac{ak}{T}\psi_{\mathrm{B}%
}=ag^{\prime}(at)+h(at).
\]
In particular, if $g$ and $h$ are compactly supported, they influence the
dynamics of the boundary system only for some finite time, after which it
returns exponentially to equilibrium.

\subsubsection{The Lamb model\label{SubSubLambModel}}

Our next example is the so called Lamb model, \cite{Lamb}, \cite{BHW}. This
time we attach a mass $m$ to the common end of the string and the spring, see
Figure \ref{Figure1} (b), keeping the rest of the assumptions. Surprisingly
enough, the effect of the string on the mass dynamics is that of standard
(instantaneous) damping. The boundary Lagrangian now contains a kinetic part%
\[
L_{\mathrm{B}}=\frac{m}{2}\left(  \mathcal{\partial}_{t}\psi_{\mathrm{B}%
}\right)  {}^{2}-\frac{k}{2}\psi_{\mathrm{B}}^{2},
\]
and the total Lagrangian is
\[
L\mathcal{(\partial}_{t}\psi_{\mathrm{D}},\mathcal{\partial}_{z}%
\psi_{\mathrm{D}},\psi_{\mathrm{B}})=\frac{1}{2}\int_{0}^{\infty}\left[
\rho(\mathcal{\partial}_{t}\psi_{\mathrm{D}})^{2}-T(\mathcal{\partial}_{z}%
\psi_{\mathrm{D}})^{2}\right]  \,\mathrm{dz}+\frac{m}{2}\left(
\mathcal{\partial}_{t}\psi_{\mathrm{B}}\right)  {}^{2}-\frac{k}{2}%
\psi_{\mathrm{B}}^{2}.
\]
Here we have again $\psi_{\mathrm{B}}(t)=$ $\psi_{\mathrm{L}}(t)=\psi
_{\mathrm{D}}(0,t)$. The DEL equation is the wave equation (\ref{waveeq}),
whereas the BEL equation is%
\begin{equation}
T\mathcal{\partial}_{z}\psi_{\mathrm{D}}(0,t)-k\psi_{\mathrm{B}}%
(t)-m\mathcal{\partial}_{tt}^{2}\psi_{\mathrm{B}}(t)=0. \label{BCLamb}%
\end{equation}
If we assume the non-radiation condition and replace, as before,
$\mathcal{\partial}_{z}\psi_{\mathrm{D}}(0,t)$ by $-\mathcal{\partial}_{t}%
\psi_{\mathrm{B}}(t)/a,$ we are left with
\[
m\mathcal{\partial}_{tt}^{2}\psi_{\mathrm{B}}+\frac{T}{a}\mathcal{\partial
}_{t}\psi_{\mathrm{B}}+k\psi_{\mathrm{B}}=0,
\]
which is the standard damped oscillator with instantaneous damping. The
previous example (\ref{LambNoMass}) can be seen as the limit case
$m\rightarrow0.$

Observe that both in this example and in the previous one the boundary system
is dissipative because of the non-radiation condition imposed on the string.
If we consider a finite string, the boundaries would not behave like
dissipative systems, since the energy would be constantly flying from the
string to the boundary and viceversa.

\subsubsection{Lamb model with friction retardation\label{SubSubLambModified}}

Next, we present an example of our approach that produces a generalization of
the Lamb model. We modify the standard Lamb model by attaching an additional
spring that connects the end of the string to the mass. This situation is
represented in Figure \ref{Figure1} (b) where the mass has been shifted by
means of a bracket to make room for the secondary springs. The introduction of
such secondary springs makes the deflection of the end of the string and that
of the point mass different, in general. According to the general approach, we
denote by $\psi_{\mathrm{D}}=\psi_{\mathrm{D}}(z,t)$ the transverse
displacement of the string. The domain and boundary Lagrangian densities are
as in the Lamb model, \ but now a non-zero interaction Lagrangian accounts for
the energy stored in the secondary springs:
\[
L_{\mathrm{INT}}\mathcal{(}\psi_{\mathrm{B}},\psi_{\mathrm{D}})=-\frac
{\widetilde{k}}{2}(\psi_{\mathrm{B}}-\psi_{\mathrm{L}})^{2},
\]
where $\widetilde{k}$ stands for their Hooke constant. The total Lagrangian is%
\[
L\mathcal{(\partial}_{t}\psi_{\mathrm{D}},\mathcal{\partial}_{z}%
\psi_{\mathrm{D}},\psi_{\mathrm{B}})=\frac{1}{2}\int_{0}^{\infty}\left[
\rho(\mathcal{\partial}_{t}\psi_{\mathrm{D}})^{2}-T(\mathcal{\partial}_{z}%
\psi_{\mathrm{D}})^{2}\right]  \,\mathrm{dz}+\frac{m}{2}\left(
\mathcal{\partial}_{t}\psi_{\mathrm{B}}\right)  {}^{2}-\frac{k}{2}%
\psi_{\mathrm{B}}^{2}-\frac{\widetilde{k}}{2}(\psi_{\mathrm{B}}-\psi
_{\mathrm{L}})^{2}%
\]
As before, the DEL equation is the wave equation for the string (\ref{waveeq}%
), whereas the IEL-BEL equations are%
\begin{gather}
T\partial_{z}\psi_{\mathrm{D}}(0,t)+\widetilde{k}\left(  \psi_{\mathrm{B}%
}(t)-\psi_{\mathrm{D}}(0,t)\right)  =0;\label{systBoundary}\\
-k\psi_{\mathrm{B}}(t)-m\partial_{tt}\psi_{\mathrm{B}}(t)-\widetilde{k}\left(
\psi_{\mathrm{B}}(t)-\psi_{\mathrm{D}}(0,t)\right)  =0.\nonumber
\end{gather}
If we assume as before that no waves arise from infinity, $\ $then
$\mathcal{\partial}_{z}\psi_{\mathrm{D}}(0,t)=-\mathcal{\partial}_{t}%
\psi_{\mathrm{D}}(0,t)/a=-\mathcal{\partial}_{t}\psi_{\mathrm{L}}(t)/a$ and
the following equation for $\psi_{\mathrm{L}}(t)$ follows%
\begin{equation}
\partial_{t}\psi_{\mathrm{L}}+\frac{a\widetilde{k}}{T}\psi_{\mathrm{L}}%
=\frac{a\widetilde{k}}{T}\psi_{\mathrm{B}}. \label{evolpsiL}%
\end{equation}
The above equation describes the evolution of the field $\psi_{\mathrm{L}}$
under the influence of the boundary. Thus, the right hand side can be
interpreted as a force exerted by the boundary on the interior, and
$\psi_{\mathrm{L}}$ as the corresponding displacement. According to the
terminology of the Linear Response Theory, the \emph{response function} of the
interior is
\begin{equation}
\phi(t)=\text{\textsf{L}}^{-1}\left[  \left(  s+\frac{a\widetilde{k}}%
{T}\right)  ^{-1}\right]  (t)=\mathrm{e}^{-\frac{a\widetilde{k}}{T}t},
\label{responsefunction}%
\end{equation}
where \textsf{L}$^{-1}$ stands for the inverse Laplace transform. Hence, the
general solution of (\ref{evolpsiL}) is given by
\begin{equation}
\psi_{\mathrm{L}}(t)=\psi_{\mathrm{L}}(0)\mathrm{e}^{-\frac{a\widetilde{k}}%
{T}t}+\frac{a\widetilde{k}}{T}\int_{0}^{t}\mathrm{e}^{-\frac{a\widetilde{k}%
}{T}(t-\tau)}\psi_{\mathrm{B}}(\tau)\,\mathrm{d}\tau. \label{ResponseInterior}%
\end{equation}
Thus, the presence of the interaction Lagrangian introduces a
history-dependent response of the interior. (\ref{ResponseInterior}) replaces
the continuity constraint $\psi_{\mathrm{L}}(t)=\psi_{\mathrm{B}}(t).$

Let us next find the evolution of $\psi_{\mathrm{B}}.$ It follows from
(\ref{ResponseInterior}) and the second equation in (\ref{systBoundary}) that
$\psi_{\mathrm{B}}$ is a solution of the integro-differential equation%
\begin{equation}
m\partial_{tt}\psi_{\mathrm{B}}+\left(  k+\widetilde{k}\right)  \psi
_{\mathrm{B}}-\widetilde{k}\psi_{\mathrm{L}}(0)\mathrm{e}^{-\frac
{a\widetilde{k}}{T}t}-\frac{a\widetilde{k}^{2}}{T}\int_{0}^{t}\mathrm{e}%
^{-\frac{a\widetilde{k}}{T}(t-\tau)}\psi_{\mathrm{B}}(\tau)\,\mathrm{d}\tau=0.
\label{GenLamb}%
\end{equation}
Integrating by parts the last term above, we get%
\[
\int_{0}^{t}\mathrm{e}^{-\frac{a\widetilde{k}}{T}(t-\tau)}\psi_{\mathrm{B}%
}(\tau)\,\mathrm{d}\tau=\frac{T}{a\widetilde{k}}\left[  \psi_{\mathrm{B}%
}(t)-\mathrm{e}^{-\frac{a\widetilde{k}}{T}t}\psi_{\mathrm{B}}(0)-\int_{0}%
^{t}\mathrm{e}^{-\frac{a\widetilde{k}}{T}(t-\tau)}\partial_{t}\psi
_{\mathrm{B}}(\tau)\,\mathrm{d}\tau\right]  ,
\]
hence (\ref{GenLamb}) becomes%
\begin{equation}
m\partial_{tt}\psi_{\mathrm{B}}+k\psi_{\mathrm{B}}+\widetilde{k}%
\psi_{\mathrm{B}}(0)\mathrm{e}^{-\frac{a\widetilde{k}}{T}t}-\widetilde{k}%
\psi_{\mathrm{L}}(0)\mathrm{e}^{-\frac{a\widetilde{k}}{T}t}+\widetilde{k}%
\int_{0}^{t}\mathrm{e}^{-\frac{a\widetilde{k}}{T}(t-\tau)}\partial_{t}%
\psi_{\mathrm{B}}(\tau)\,\mathrm{d}\tau=0. \label{GenLamb1}%
\end{equation}
If we assume $\psi_{\mathrm{B}}(0)=$ $\psi_{\mathrm{L}}(0),$ we are left with%
\begin{equation}
m\partial_{tt}\psi_{\mathrm{B}}+k\psi_{\mathrm{B}}+\widetilde{k}\int_{0}%
^{t}\mathrm{e}^{-\frac{a\widetilde{k}}{T}(t-\tau)}\partial_{t}\psi
_{\mathrm{B}}(\tau)\,\mathrm{d}\tau=0, \label{GenLamb2}%
\end{equation}
which models a damped oscillator with linear damping depending on all the
evolution from $\tau=0$ to $\tau=t.$ Observe that the convolution kernel is a
decaying exponential, meaning that the values of velocity close to $\tau=t$
are more relevant in the averaging than those, close to $\tau=0.$

Observe that in the case $\widetilde{k}=0,$ we recover the free harmonic
oscillator, $m\partial_{tt}\psi_{\mathrm{B}}+k\psi_{\mathrm{B}}=0.$ This is
natural, since in this case there is no interaction between the string and the oscillator.

If $\widetilde{k}\rightarrow\infty,$ we expect to recover the standard Lamb
model, since in that case the string is attached rigidly to the mass. Our
system is simple enough to study this limit explicitly. For $\varepsilon>0$,
the dissipative term in \ref{GenLamb2} can be split as%
\[
\widetilde{k}\int_{0}^{t}\mathrm{e}^{-\frac{a\widetilde{k}}{T}(t-\tau
)}\partial_{t}\psi_{\mathrm{B}}(\tau)\,\mathrm{d}\tau=\widetilde{k}\int
_{0}^{t-\varepsilon}\mathrm{e}^{-\frac{a\widetilde{k}}{T}(t-\tau)}\partial
_{t}\psi_{\mathrm{B}}(\tau)\,\mathrm{d}\tau+\widetilde{k}\int_{t-\varepsilon
}^{t}\mathrm{e}^{-\frac{a\widetilde{k}}{T}(t-\tau)}\partial_{t}\psi
_{\mathrm{B}}(\tau)\,\mathrm{d}\tau.
\]
The first integral above vanishes as $\widetilde{k}\rightarrow\infty,$ for any
$\varepsilon>0$ fixed. Indeed,%
\[
\widetilde{k}\int_{0}^{t-\varepsilon}\mathrm{e}^{-\frac{a\widetilde{k}}%
{T}(t-\tau)}\partial_{t}\psi_{\mathrm{B}}(\tau)\,\mathrm{d}\tau\leq
\widetilde{k}\mathrm{e}^{-\frac{a\widetilde{k}}{T}\varepsilon}\left[
\psi_{\mathrm{B}}(t-\varepsilon)-\psi_{\mathrm{B}}(0)\right]  \rightarrow
0\qquad\mathrm{as}\text{ \ }\widetilde{k}\rightarrow\infty
\]
uniformly on bounded intervals of $t.$ As to the second integral, observe
that, for $b>0,$%
\[
A\mathrm{e}^{-bAs}\rightarrow\frac{1}{b}\delta(s),\qquad\mathrm{as}\text{
\ }A\rightarrow\infty
\]
in the sense of distributions. Therefore,%
\[
\widetilde{k}\int_{t-\varepsilon}^{t}\mathrm{e}^{-\frac{a\widetilde{k}}%
{T}(t-\tau)}\partial_{t}\psi_{\mathrm{B}}(\tau)\,\mathrm{d}\tau=\widetilde
{k}\int_{0}^{\varepsilon}\mathrm{e}^{-\frac{a\widetilde{k}}{T}s}\partial
_{t}\psi_{\mathrm{B}}(t-s)\,\mathrm{d}\tau\rightarrow\frac{T}{a}\partial
_{t}\psi_{\mathrm{B}}(t)
\]
and, finally, (\ref{GenLamb2}) becomes%
\begin{equation}
m\partial_{tt}\psi_{\mathrm{B}}+\frac{T}{a}\partial_{t}\psi_{\mathrm{B}}%
+k\psi_{\mathrm{B}}=0, \label{StandardLamb}%
\end{equation}
which is precisely the standard Lamb model.

The dissipative force in (\ref{GenLamb2}), which depends on the past values of
the velocity, is a particular instance of the so called \emph{retarded
frictional forces, }as opposed to instantaneous friction present in
(\ref{StandardLamb}).

It is clear that we can add springs and masses to our system. This procedure
leads to more complex response functions and dissipative forces in the
equation for $\psi_{\mathrm{B}}.$Thus for the system represented in Figure
\ref{Figure1} (d), in which a mass has been added at the end of the string,
the friction term in the equation for $\psi_{\mathrm{B}}$ has the form
\begin{equation}
c_{1}\int_{0}^{t}\mathrm{e}^{-c_{2}(t-\tau)}\sin\left(  c_{3}\tau\right)
\partial_{t}\psi_{\mathrm{B}}(\tau)\,\mathrm{d}\tau\label{ComplexResponse}%
\end{equation}
for positive constants $c_{i}$ depending on the parameters of the system.

\subsection{Multi-transmission lines\label{SubSectionMTL}}

Next, we apply the above general setting to a (lossless) multi-transmission
line, MTL. Suppose that we have $k+1$ conductors, one of them being grounded,
say the $(k+1)$-th. \ We denote by $V(z,t)=\left\{  V^{\ell}(z,t)\right\}
_{\ell=1\ldots k}$ the vector-column of voltages \ on \ the first $k$
conductors with respect to the ground and \ by $I(z,t)=\left\{  I^{\ell
}(z,t)\right\}  _{\ell=1\ldots k}$ the vector-column of currents flowing on
them. The dynamics of the system is described by the telegrapher's equations
\begin{equation}
\partial_{t}V=-\mathbf{C}^{-1}\partial_{z}I,\qquad\partial_{t}I=-\mathbf{L}%
^{-1}\partial_{z}V, \label{tranbe2}%
\end{equation}
where $\mathbf{C}$ (respectively $\mathbf{L})$ \ are the matrices of mutual
capacity (respectively mutual inductance) of the conductors, see \cite{MM}. A
Lagrangian setup can be introduced by defining generalized coordinates-fields
as follows:%
\[
\psi_{\mathrm{D}}^{\ell}(z,t)=(\psi_{\mathrm{D}}^{\ell}(z,t))_{\ell=1\ldots
k},\text{ \qquad}\psi_{\mathrm{D}}^{\ell}(z,t)=%
{\displaystyle\int\limits^{t}}
I^{\ell}(z,s)ds.
\]
The corresponding Lagrangian density is%
\begin{equation}
L_{\mathrm{D}}=\frac{1}{2}\left(  \partial_{t}\psi_{\mathrm{D}}^{\ell
},\mathbf{L}\partial_{t}\psi_{\mathrm{D}}^{\ell}\right)  -\frac{1}{2}\left(
\partial_{z}\psi_{\mathrm{D}}^{\ell},\mathbf{C}^{-1}\partial_{z}%
\psi_{\mathrm{D}}^{\ell}\right)  , \label{mtraneq1}%
\end{equation}
where $\left(  \cdot,\cdot\right)  $ denotes the standard scalar product in $%
\mathbb{C}
^{n}$.

Suppose we have a semi-infinite transmission line (we take $k=1$ for
simplicity, the generalization is straightforward). The regime at the end of
the transmission line can be imposed by attaching another system, so called
"load" in circuit theory. These loads play the role of boundary systems.
Linear passive loads are uniquely characterized by their impedance, which
relates the voltage between the terminals to the current flowing in the load,
in the Fourier-Laplace domain, see subsection \ref{SubSectResponseTheory} in
Appendix. Thus for example an RLC load has impedance $Z_{\mathrm{RLC}%
}=R+Ls+1/Cs$ . Observe that this formula is completely analogous to the
impedance of a damped oscillator, formula (\ref{ImpedanceDampedOscillator}),
via identifications mass $\sim$ inductance, damping constant $\sim$ resistance
and Hooke constant $\sim$ inverse capacity.

In agreement with our general one-dimensional setting, we introduce a boundary
field $\psi_{\mathrm{B}}$ describing the load. A general LC \ load corresponds
to a boundary Lagrangian of the form%
\[
L_{\mathrm{B}}(\psi_{\mathrm{B}},\partial_{t}\psi_{\mathrm{B}})=\frac
{\widetilde{\mathbf{L}}}{2}\left(  \partial_{t}\psi_{\mathrm{B}}\right)
^{2}-\frac{1}{2\widetilde{\mathbf{C}}}\psi_{\mathrm{B}}^{2},
\]
where $\widetilde{\mathbf{L}}$ \ and $\widetilde{\mathbf{C}}$ are the
inductance and capacity of the LC load (observe the analogy with the
mass-spring system). In order to enforce the continuity condition%
\[
\psi_{\mathrm{B}}(t)=\psi_{\mathrm{L}}(t)=\psi_{\mathrm{D}}(0,t),\qquad t>0
\]
we can proceed as in the example of Subsection \ref{SubSubLambModified}.
First, we do not assume any constraint and introduce an interaction Lagrangian
of the form%
\[
L_{\mathrm{INT}}\mathcal{(}\psi_{\mathrm{B}},\psi_{\mathrm{D}})=-\frac{A}%
{2}(\psi_{\mathrm{B}}-\psi_{\mathrm{L}})^{2},\qquad A>0.
\]
Then, we derive the corresponding DEL and BEL equations and, finally, we let
$A\rightarrow\infty.$ An interaction Lagrangian of the above form can be
interpreted as the insertion of a capacitor of capacity $1/A$ between the TL
and the load. As $A\rightarrow\infty,$ its capacity vanishes and the
continuity condition is achieved.

Formally, our Lagrangian system is the same as the one in Subsection
\ref{SubSubLambModified}, if we make the identifications:%
\[
\mathbf{L}\longleftrightarrow\rho;\quad\mathbf{C}^{-1}\longleftrightarrow
T;\quad\widetilde{\mathbf{L}}\longleftrightarrow m;\quad\widetilde{\mathbf{C}%
}^{-1}\longleftrightarrow k;\quad A\longleftrightarrow\widetilde{k}.
\]
Therefore, the resulting EL system is%
\begin{align}
\text{(DEL) \ \ \ \ \ \ \ \ \ \ \ \ \ \ }\mathbf{L}\partial_{tt}^{2}%
\psi_{\mathrm{D}}-\mathbf{C}^{-1}\partial_{zz}^{2}\psi_{\mathrm{D}}  &
=0,\qquad z>0,t>0;\label{MultiLineSystem}\\
\text{(IEL) \ \ \ \ \ \ \ \ \ \ \ }\mathbf{C}^{-1}\partial_{z}\psi
_{\mathrm{D}}(0,t)+A(\psi_{\mathrm{B}}(t)-\psi_{\mathrm{D}}(0,t))  &
=0\qquad,t>0;\nonumber\\
\text{(BEL) \ \ \ \ \ \ \ \ \ \ \ }-\widetilde{\mathbf{C}}^{-1}\psi
_{\mathrm{B}}(t)-\widetilde{\mathbf{L}}\partial_{tt}\psi_{\mathrm{B}%
}(t)-A(\psi_{\mathrm{B}}(t)-\psi_{\mathrm{D}}(0,t))  &  =0\qquad,t>0.\nonumber
\end{align}
The computations in Subsection \ref{SubSubLambModified} show that if we assume
that no wave is arising from infinity on the TL, after taking the limit
$A\rightarrow\infty,$ we get the boundary evolution
\[
\widetilde{\mathbf{L}}\partial_{tt}\psi_{\mathrm{B}}+\sqrt{\mathbf{LC}^{-1}%
}\partial_{t}\psi_{\mathrm{B}}+\widetilde{\mathbf{C}}^{-1}\psi_{\mathrm{B}%
}=0.
\]
We conclude that the TL acts on the load as instantaneous damping due to an
effective resistance $R=\sqrt{\mathbf{LC}^{-1}},$ so called characteristic
impedance of the TL, see \cite{MM}. For finite $A,$ the effective damping is
not instantaneous. In this case, the impedance of the load is
frequency-dependent. It is worth observing in this example that the effect of
the boundary on the interior is reflected as a boundary condition (not
affecting the master equation of the continuous system) whereas the reciprocal
effect of the continuous system on the boundary modifies the equation.

If the attached load contains a voltage source $V(t),$ it can be incorporated
in the l.h.s. of the BEL equations.

\subsection{The oscillating membrane\label{SubSectMembrane}}

In this section we apply our general framework to a sequence of
two-dimensional examples related to the trampoline described in the
Introduction. Aimed at describing small vertical oscillations of the membrane
(jumping mat) and the frame we make the following general assumptions: a) the
membrane is two-dimensional and flexible with surface mass density $\sigma$;
b) the frame is a rigid one-dimensional ring with linear mass density
$\lambda;$c) the frame is elastically supported by means of springy poles with
distributed Hooke constant per unit length $k;$e) lateral displacements are
neglected and f) oscillations are small enough to justify linearization around
the equilibrium. Accordingly, we only keep quadratic terms in the energies.
Note that in real trampolines, the mat is attached to the frame by means of
horizontal springs which keep it taut. Under small vertical displacements of
the mat, the elastic energy stored in those springs is constant in linear
approximation and plays no role in the energy balance.

As in the one dimensional case, we present a sequence of increasingly complex boundaries.

We start by assuming that the ring is massless $(\lambda=0).$This case
corresponds to the one-dimensional example from subsection
\ref{SubSubSimpleSpring}.

Denote by $\Omega$ the plane circular region on which the membrane is
projected at all times and by $\Gamma$ its boundary, that is, the projection
of the ring. Let $\psi_{\mathrm{D}}(x,y,t)$ be the vertical deflection of the
point of the membrane projected on $(x,y)\in\Omega$ at time $t,$ where $(x,y)$
are Cartesian coordinates. The boundary is the frame-poles system. Denoting by
$s$ the arc length parameter along the ring, the vertical deflection of the
ring is the boundary field $\psi_{\mathrm{B}}(s,t).$(the length is measured,
strictly speaking, along the projection of the ring, not along the ring, but
this difference is irrelevant in the linear approximation).

The kinetic energy of the membrane is, as usual,%
\[
K_{\mathrm{M}}=\frac{\sigma}{2}%
{\displaystyle\int\limits_{\Omega}}
\left(  \partial_{t}\psi_{\mathrm{D}}\right)  ^{2}\,\mathrm{dxdy.}%
\]
In the quadratic approximation, the elastic energy stored in the membrane is%
\[
U_{\mathrm{M}}=\frac{T}{2}%
{\displaystyle\int\limits_{\Omega}}
\left\vert \nabla\psi_{\mathrm{D}}\right\vert ^{2}\,\mathrm{dxdy,}%
\]
where $T$ is the tension in the membrane, that is, the elastic force per unit
length perpendicular to a linear element, see \cite{FG}. Such tension is a
constant in the small oscillations approximation. The two-dimensional domain
Lagrangian density is%
\[
L_{\mathrm{D}}=\frac{\sigma}{2}\left(  \partial_{t}\psi_{\mathrm{D}}\right)
^{2}-\frac{T}{2}\left\vert \nabla\psi_{\mathrm{D}}\right\vert ^{2}.
\]
The elastic energy stored in the support is
\begin{equation}
U_{\mathrm{SUPP}}=\frac{1}{2}%
{\displaystyle\int\limits_{\Gamma}}
k\psi_{\mathrm{B}}^{2}(s,t)\,\mathrm{ds.} \label{EnergySpringsMem/Ring}%
\end{equation}
and the corresponding linear Lagrangian density is%
\[
L_{\mathrm{B}}=-\frac{1}{2}k\psi_{\mathrm{B}}^{2}(s,t)\,
\]
The action is given by
\[
S[\psi_{\mathrm{D}},\psi_{\mathrm{B}}]=\int_{t_{0}}^{t_{1}}\left[
{\displaystyle\int\limits_{\Omega}}
L_{\mathrm{D}}\mathrm{dxdy+}%
{\displaystyle\int\limits_{\Gamma}}
L_{\mathrm{B}}\mathrm{ds}\right]  \,\mathrm{dt}.
\]
Following the general procedure described in section
\ref{SubSectMultipleDimensions}, and taking into account the continuity
condition
\[
\psi_{\mathrm{D}}(x(s),y(s),t)=\psi_{\mathrm{L}}(s,t)=\psi_{\mathrm{B}}(s,t)
\]
we get the following equations%
\begin{gather}
(\mathrm{DEL})\qquad\qquad\partial_{tt}^{2}\psi_{\mathrm{D}}=a^{2}\Delta
\psi_{\mathrm{D}}\qquad\text{on }\Omega,\quad a^{2}=\frac{T}{\sigma
};\label{SimpleMembrane}\\
(\mathrm{BEL})\qquad\qquad k\psi_{\mathrm{B}}+T\partial_{\mathrm{n}}%
\psi_{\mathrm{B}}=0\qquad\text{on }\Gamma,\nonumber
\end{gather}
where $\Delta$ denotes the Laplace operator, $\Delta f=\partial_{xx}%
^{2}f+\partial_{yy}^{2}f$ in Cartesian coordinates, and $\partial_{\mathrm{n}%
}$ stands for the exterior normal derivative. Observe that since we are taking
the arc length as a parameter on the boundary, then%
\[
g=\left\vert \frac{d\mathbf{r}}{ds}\right\vert =1,
\]
where $\mathbf{r:\,}s\in\left[  0,l\right]  \rightarrow\mathbb{E}^{2}$ is the
parametric embedding of the boundary. Consequently, $dg/ds=0$ and covariant
derivatives on the boundary reduce to standard derivatives with respect to
$s.$We easily recognize in the $(\mathrm{BEL})$ equation in
(\ref{SimpleMembrane}) the two-dimensional Robin boundary conditions. If we
let $k\rightarrow\infty,$ we are left with the standard "clamped" membrane,
with Dirichlet boundary conditions%
\[
\psi_{\mathrm{D}}=0\qquad\text{on }\Gamma.
\]
Equations (\ref{SimpleMembrane}) can be easily rewritten in polar coordinates.

Let us next include the effect of a heavy frame $(\lambda\neq0)$. Now we have
a new contribution to the energy of the system: the kinetic energy of the
ring, given by%
\[
K_{\mathrm{R}}=\frac{1}{2}%
{\displaystyle\int\limits_{\Gamma}}
\lambda\left(  \partial_{t}\psi_{\mathrm{B}}\right)  ^{2}(s,t)\,\mathrm{ds,}%
\]
This situation clearly falls into the general framework described in Section
\ref{SubSectMultipleDimensions} with%
\[
L_{\mathrm{D}}=K_{\mathrm{M}}-U_{\mathrm{M}};\qquad L_{\mathrm{B}%
}=K_{\mathrm{R}}-U_{\mathrm{SUPP}};\qquad L_{\mathrm{INT}}=0.
\]
(actually, as we remarked in Subsection \ref{MainResult1D}, one should first
consider a nontrivial interaction and then take limits). According to general
formulas (\ref{MDEL})-(\ref{MBEL}), the equation for the interior system is
still the two-dimensional wave equation in (\ref{SimpleMembrane}), whereas the
BEL equations reduce to%
\begin{equation}
\lambda\partial_{tt}^{2}\psi_{\mathrm{B}}+k\psi_{\mathrm{B}}+T\partial
_{\mathrm{n}}\psi_{\mathrm{D}}=0\qquad\text{on }\Gamma.
\label{BELRigidMembrane}%
\end{equation}

We can conceive more sophisticated "boundaries". For example, if the membrane
is suspended by means of vertical springs with distributed (per unit length)
Hooke constant $\widetilde{k}$, the elastic energy stored in the springs is
given by%
\begin{equation}
U_{\mathrm{Spring}}=\frac{1}{2}%
{\displaystyle\int\limits_{\Gamma}}
\widetilde{k}\left[  \psi_{\mathrm{D}}(x(s),y(s),t)\,-\psi_{\mathrm{B}%
}(s)\right]  ^{2}\mathrm{ds.}%
\end{equation}
In this case we have%
\[
L_{\mathrm{D}}=K_{\mathrm{M}}-U_{\mathrm{M}};\qquad L_{\mathrm{B}%
}=K_{\mathrm{R}}-U_{\mathrm{SUPP}};\qquad L_{\mathrm{INT}}=-U_{\mathrm{Spring}%
}.
\]
Using the general formulas (\ref{MDEL})-(\ref{MBEL}), we get the following
IEL-BEL equations:%
\begin{align*}
-T\partial_{\mathrm{n}}\psi_{\mathrm{D}}+\widetilde{k}\left[  \psi
_{\mathrm{D}}(x(s),y(s),t)-\psi_{\mathrm{B}}(s,t)\right]   &  =0;\\
-k\psi_{\mathrm{B}}(s,t)-\lambda\partial_{tt}^{2}\psi_{\mathrm{B}%
}(s,t)-\widetilde{k}\left[  \psi_{\mathrm{D}}(x(s),y(s),t)\,-\psi_{\mathrm{B}%
}(s)\right]   &  =0.
\end{align*}

In the limit \ $\widetilde{k}\rightarrow\infty$ we recover the "rigid"
constraint and the above equations reduce to (\ref{BELRigidMembrane}).

More complex situations can be considered within our framework. For example,
we can allow the frame to be flexible. In such case, we have to add to the
boundary Lagrangian the potential energy associated to flexion. Such
Lagrangian would involve the tangential derivative $\partial\psi_{\mathrm{B}%
}/\partial s.$ Nonlinear effects associated to large vertical displacement can
also be considered.

\section{Connection to conservative
extensions\label{SectConservativeExtensions}}

According to \cite{FS}, very general dissipative and dispersive systems of the
form
\begin{equation}
m\partial_{t}v=-\mathrm{i}Av-\int_{0}^{\infty}a(\tau)v(t-\tau)\,\mathrm{d\tau
}+f(t) \label{gensystem}%
\end{equation}
can be extended to conservative systems by means of attaching a convenient
system of strings which can be interpreted as hidden degrees of freedom.

Looking at our one-dimensional examples from subsection
\ref{SubSectOneDimExamples} from the "boundary" point of view provides
explicit examples of such conservative extensions. Indeed, as we noted in
subsection \ref{SubSectOneDimExamples}, the boundary systems behave as a
dissipative systems. The rest of the original conservative system can be thus
interpreted as the one associated to the hidden degrees of freedom. It is
important that those degrees of freedom be at complete rest before $t=0$ and
only subjected to the forces arising from their link to the dissipative
system. In our examples, such condition was enforced by assuming trivial
initial data and the non-radiation condition at infinity.

The example of the Lamb model, discussed in subsection \ref{SubSubLambModel},
was already pointed out in \cite{FS}. In this case, frictional forces are
instantaneous, according to the terminology introduced in Subsection
\ref{SubSectResponseTheory} \ of the Appendix. The hidden degrees of freedom
are here provided by the string, that takes the energy away from the boundary.

Our one-dimensional structured boundaries are a source of arbitrarily complex
dissipative and dispersive systems and their corresponding conservative
extensions. Thus, the example in Subsection \ref{SubSubLambModified} provides
a concrete conservative extension for the dispersive system (\ref{GenLamb2}),
which can be easily recast in the form (\ref{gensystem}), see subsection
\ref{SubSectResponseTheory} in the Appendix. In this case we have a
\textit{genuine dispersive system} with exponential response function. More
sophisticated boundaries give rise to other response functions like that in
(\ref{ComplexResponse}).

\section{Appendix\label{SectAppendix}}

\subsection{On linear response theory\label{SubSectResponseTheory}}

Linear response theory deals with linear relations between applied forces and
the corresponding displacements or currents. Its main concepts are discussed
in \cite{Ku}, and a concrete mathematical framework has been given in
\cite{FS}. We briefly recall here the main facts and terminology, following
the above references.

Very general linear dynamical systems are governed by equations of the form
(\ref{gensystem}), where $v(t)$ is the state of the system, belonging to some
Hilbert space $H,$ $m$ is a positive mass operator, $A:D(A)\rightarrow H$ is a
self-adjoint operator, $a(t),t\geq0$ is a family of operators and $f$ is an
$H-$ valued function, defined on $[0,\infty).$ The physical interpretation is
the following: the terms%
\begin{equation}
-iAv-\int_{0}^{\infty}a(\tau)v(t-\tau)\,\mathrm{d\tau} \label{selfforce}%
\end{equation}
represent the self-force. The term $-\mathrm{i}Av$ corresponds to conservative
forces, such as (linear) elastic forces. The work done by those forces
$-\mathrm{i}Av$ is transformed into kinetic energy of the system, as a
consequence of the unitarity of the group $\mathrm{e}^{-\mathrm{i}At}.$ The
second term in (\ref{selfforce}) corresponds to frictional forces of a very
general nature. Indeed, in many cases $a(\tau)$ has the structure%
\[
a(\tau)=\alpha_{\infty}\delta(\tau)+\alpha(\tau)
\]
where $\alpha_{\infty}$ is some fixed self-adjoint operator and $\alpha(\tau)$
is continuous for $\tau\geq0.$ The corresponding friction force in
(\ref{selfforce}) has the form%
\begin{equation}
f_{\mathrm{fric}}(t)=\alpha_{\infty}v(t)+\int_{0}^{\infty}\alpha(\tau
)v(t-\tau)\,\mathrm{d\tau.} \label{RetardedFrictionalForce}%
\end{equation}
The term $\alpha_{\infty}\delta(\tau)$ corresponds to \textsl{instantaneous
friction}, depending only on the current state of the system, whereas the
second term describes \textsl{retarded friction}, dependent upon the history
of the system prior to time $t$ (causality).

In order that $a(\tau)$ represent real dissipation, sign restrictions should
be imposed. The total work of the frictional forces per unit of time can be
written as%
\[
W_{\mathrm{fric}}=-\frac{1}{2}\int_{-\infty}^{\infty}\int_{-\infty}^{\infty
}\left\langle v(t),a_{e}(\tau)v(t-\tau)\,\right\rangle _{H}\,\mathrm{d\tau
\,dt},
\]
where the operator $a_{e}(\tau)$ has been introduced to get a more symmetric
expression and is related to $a(\tau)$ by the formula%
\begin{equation}
a_{e}(\tau)=2\alpha_{\infty}\delta(t)+\left\{
\begin{array}
[c]{c}%
\alpha(\tau),\qquad\tau>0\\
\operatorname{Re}\alpha(0^{+}),\qquad\tau=0\\
\alpha^{\ast}(-\tau),\qquad\tau<0
\end{array}
\right.  . \label{ae}%
\end{equation}
Condition $W_{\mathrm{fric}}\leq0$ means that $a_{e}(\tau)$ is a positive
definite function, a fact that plays a crucial role in the construction of a
conservative extension of the system given in \cite{FS}.

The response of the system to the applied external force is the mapping
$\ f(t)\rightarrow v(t),$ where we assume that the system was at rest for
$t\leq0$. It is often more convenient to describe it in the frequency
domain,\textit{ i.e.} in terms of the Laplace-Fourier transforms of $f$ and
$v.$ We define the Laplace-Fourier transform of an $H-$ valued function
$u(t),$ defined for $t\geq0,$ as%
\[
\widehat{u}(\zeta)=\int_{0}^{\infty}u(t)\mathrm{e}^{\mathrm{i}\zeta
t}\,\mathrm{dt,\qquad}\zeta=\omega+\mathrm{i}\eta\mathrm{.}%
\]
$\widehat{u}$ is well defined in a half-plane of the form $\eta>\eta_{0},$
where $\eta_{0}$ depends on the exponential class of $u.$ In particular, if
$u$ is bounded for $t\geq0,$ $\widehat{u}$ is well defined in the upper
half-plane $\eta>0$ and is a holomorphic function of $\zeta$ there.

Formally applying the Laplace-Fourier transform to the equation
(\ref{gensystem}), we obtain%
\[
\zeta m\widehat{v}(\zeta)=[A-i\widehat{a}(\zeta)]\widehat{v}(\zeta
)+\mathrm{i}\widehat{f}(\zeta),
\]
which leads to the representation%
\begin{equation}
\widehat{v}(\zeta)=\mathfrak{A}(\zeta)\widehat{f}(\zeta),
\label{admittancerepresentation}%
\end{equation}
where the \textsl{admittance operator} is defined as%
\begin{equation}
\mathfrak{A}(\zeta)=\mathrm{i}[\zeta m-A+\mathrm{i}\widehat{a}(\zeta)]^{-1}.
\label{admittancedefinition}%
\end{equation}
The admittance representation (\ref{admittancerepresentation}) furnishes a
complete description of the original system. The inverse operator%
\[
\mathcal{Z}(\zeta)=\left[  \mathfrak{A}(\zeta)\right]  ^{-1}=-\mathrm{i}[\zeta
m-A+\mathrm{i}\widehat{a}(\zeta)]
\]
is called \textsl{impedance operator}.

Positive definiteness of $a_{e}$ is reflected on its Fourier-Laplace
transform. Indeed, it is equivalent to the condition%
\[
\operatorname{Im}\widehat{a}(\zeta)\geq0\qquad\text{for }\operatorname{Im}%
\zeta>0,
\]
which is in turn equivalent to the fact that the the complex function%
\[
f_{v}(\zeta):=\left\langle v,\mathrm{i}\widehat{a}(\zeta)v\right\rangle
\]
has non-negative imaginary part for $\operatorname{Im}\zeta>0$ and any fixed
$v\in H.$ This property, together with suitable boundedness assumptions on
$\alpha_{\infty},\alpha(\tau),$ allows for the following representation of
$f_{v}(\zeta)$ (via Nevanlinna Theorem)
\[
f_{v}(\zeta)=\int_{-\infty}^{\infty}\frac{dN_{v}(\sigma)}{\sigma-\zeta}%
\]
for some nondecreasing, right continuous and bounded function $N_{v}.$ See
\cite{FS} for a detailed account on these issues.

(\ref{admittancerepresentation}) is a very \ general representation. Many
important real systems evolve in $H=%
\mathbb{C}
.$ In this case, $\mathfrak{A}(\zeta)$ is a complex function,
(\ref{admittancerepresentation}) amounts to multiplication by the so called
\textsl{complex admittance }$\widehat{\mathfrak{\chi}}(\zeta):$\textsl{ }%
\[
\widehat{v}(\zeta)=\widehat{\mathfrak{\chi}}(\zeta)\widehat{f}(\zeta),
\]
and its time domain counterpart has the form%
\begin{equation}
v(t)=\chi^{\infty}f(t)+\int_{0}^{\infty}\Phi(\mathrm{\tau})f(t-\mathrm{\tau
})\,\mathrm{d\tau}=\chi^{\infty}f(t)+\int_{-\infty}^{t}\Phi
(t-\mathrm{\varsigma})f(\mathrm{\varsigma})\,\mathrm{d\varsigma,}
\label{ResponseFormula}%
\end{equation}
for some complex function $\Phi$, called the \textsl{response function}. The
terminology is justified by the fact that the solution corresponding to a unit
step force applied at $t=t_{1},$ $f(t)=\theta(t-t_{1}),$ is precisely
$v(t)=\Phi(t-t_{1})\theta(t-t_{1}),$ where $\theta$ is the Heaviside step
function. On the other hand, the constant $\chi^{\infty}$ is the response of
the system to a unit impulse at $t=0,$ $f(t)=\delta(t)$ and corresponds to the
constant part of the complex admittance. Whenever the admittance operator is
not a constant, $\Phi\neq0.$

The relation between the complex admittance and the response function is given
by the Fourier-Laplace transform%
\[
\widehat{\chi}(\zeta)=\chi^{\infty}+\int_{0}^{\infty}\Phi(\tau)\mathrm{e}%
^{\mathrm{i}\zeta\tau}\,\mathrm{d\tau.}%
\]
$\widehat{\chi}(\zeta)$ or, equivalently, the pair $\left(  \chi^{\infty}%
,\Phi(t)\right)  $ completely describe the behavior of the system.\medskip

\subsubsection{Examples}

1) The ODE for the standard damped oscillator:%
\begin{equation}
\partial_{tt}u+2a\partial_{t}u+ku=g(t);\qquad a,k>0 \label{dampedoscillator}%
\end{equation}
(we take unit mass for simplicity) can be put in the form(\ref{gensystem}) by
choosing $v=(v_{1},v_{2})\in%
\mathbb{C}
^{2}=H$ with
\[
v_{1}=\sqrt{k}u;\qquad v_{2}=u_{t}.
\]
Indeed, (\ref{dampedoscillator}) is equivalent to (\ref{gensystem}) with
$m=Id,$
\[
A=\left(
\begin{array}
[c]{cc}%
0 & \mathrm{i}\sqrt{k}\\
-\mathrm{i}\sqrt{k} & 0
\end{array}
\right)  ;\qquad\alpha_{\infty}=\left(
\begin{array}
[c]{cc}%
0 & \mathrm{0}\\
0 & 2a
\end{array}
\right)  ,\quad\alpha(\tau)=0;\qquad f(t)=\left(
\begin{array}
[c]{c}%
0\\
g(t)
\end{array}
\right)  .
\]
Thus in this case we have instantaneous damping. Clearly $\widehat{a}%
=\alpha_{\infty}$ and we can compute the impedance operator explicitly:%
\[
\mathcal{Z}(\zeta)=-\mathrm{i}[\zeta m-A+\mathrm{i}\widehat{a}(\zeta
)]=-\mathrm{i}\left(
\begin{array}
[c]{cc}%
\zeta & -\mathrm{i}\sqrt{k}\\
\mathrm{i}\sqrt{k} & \zeta+2a\mathrm{i}%
\end{array}
\right)  .
\]
Usually (for example, in the Circuit theory context) impedance is understood
as the ratio%
\[
Z=\frac{\mathsf{L}[g]}{\mathsf{L}[u_{t}]},
\]
where $\mathsf{L}$ represents the standard Laplace transform. We can easily
recover $Z$ from $\mathcal{Z}$ . Indeed, if we only consider the second
component in the relation%
\[
\widehat{f}(\zeta)=\mathcal{Z}(\zeta)\widehat{v}(s),
\]
we get
\[
\widehat{g}=\mathrm{i}\sqrt{k}\widehat{v_{1}}+(\zeta+2a\mathrm{i)}%
\widehat{v_{2}}.
\]
But $\widehat{v_{1}}=\sqrt{k}\widehat{u}=\frac{\sqrt{k}}{\mathrm{i}\zeta
}\widehat{v_{2}}.$ Therefore,
\[
\widehat{g}=\left[  \frac{\mathrm{i}k}{\zeta}+(-\mathrm{i}\zeta+2a\mathrm{)}%
\right]  \widehat{v_{2}}=\left[  \frac{k}{s}+2a+s\right]  \widehat{v_{2}%
};\qquad\zeta=\mathrm{i}s
\]
and%
\begin{equation}
Z(s)=\frac{\mathsf{L}[g](s)}{\mathsf{L}[u_{t}](s)}=\frac{\widehat{g}(\zeta
)}{\widehat{v_{2}}(\zeta)}=\frac{k}{s}+2a+s, \label{ImpedanceDampedOscillator}%
\end{equation}
which is the usual expression for the impedance of a damped oscillator.

2) The same procedure, applied to the system from Subsection
(\ref{SubSubLambModified}) with unit mass, produces a first order system of
the form(\ref{gensystem}) with%
\[
A=\left(
\begin{array}
[c]{cc}%
0 & \mathrm{i}\sqrt{k}\\
-\mathrm{i}\sqrt{k} & 0
\end{array}
\right)  ;\qquad\alpha_{\infty}=0,\quad\alpha(\tau)=\left(
\begin{array}
[c]{cc}%
0 & \mathrm{0}\\
0 & -\widetilde{k}\mathrm{e}^{-\frac{a\widetilde{k}}{T}\tau}%
\end{array}
\right)  ;\qquad f(t)=\left(
\begin{array}
[c]{c}%
0\\
g(t)
\end{array}
\right)  .
\]
and, therefore%
\[
\mathcal{Z}(\zeta)=-\mathrm{i}[\zeta m-A+\mathrm{i}\widehat{a}(\zeta
)]=-\mathrm{i}\left(
\begin{array}
[c]{cc}%
\zeta & -\mathrm{i}\sqrt{k}\\
\mathrm{i}\sqrt{k} & \zeta-\mathrm{i}\widetilde{k}\mathrm{e}^{-\frac
{a\widetilde{k}}{T}\tau}%
\end{array}
\right)  .
\]

\subsection{On the Divergence Theorem in a Riemannian
manifold\label{SubSectDG}}

In the formulation of the Euler-Lagrange equations for systems with
boundaries, a crucial step is that of integrating by parts in space and time
the action associated to the main system. In the standard approach, one
assumes that the space is flat and Euclidean coordinates are used, hence the
divergence of a vector $v=(v_{x},v_{y},v_{z})$ is given by the sum of the
partial derivatives.%
\[
\operatorname{div}v=\frac{\partial v_{x}}{\partial x}+\frac{\partial v_{y}%
}{\partial y}+\frac{\partial v_{z}}{\partial z}%
\]
Since in our approach the boundary is a system in its own right, we have to
perform the same task on possibly curved manifolds of the Euclidean space. In
such case, the divergence of a tangent vector field is not anymore given by
the sum of "partial derivatives". Instead, one has to resort to covariant
derivatives and form the corresponding combination. \ Importantly, only when
one uses the legitimate (covariant) divergence, the divergence theorem extends
to the case of curved manifolds.

In this Appendix, we gather the main concepts and formulas needed to deal with
curved manifolds. We are mainly interested in the case of manifolds, embedded
in the Euclidean space $\mathbb{E}^{n}$, with induced metrics and so called
"compatible" or Levi-Civita connection.

\subsubsection{Covariant derivatives}

On a general differentiable manifold $M$, in order to define a concept of
derivative of a tensor field (other than a scalar), we must be able to compare
the values of the field at different points. There is no "natural" way to do
this and an additional, independent structure is required. A
\textsl{connection} is given by a system of functions%
\[
\Gamma_{\,\,j,k}^{i}(x)
\]
on the manifold, so called Christoffel symbols. If coordinates on $M$ are
changed, $(x^{s})\rightarrow(x^{s\prime}),$ the Christoffel symbols must
change according to the following, non tensorial law%
\[
\Gamma_{\,\,p^{\prime},q^{\prime}}^{k^{\prime}}=\sum_{k,p,q}\left(
\Gamma_{\,\,p,q}^{k}\frac{\partial x^{k^{\prime}}}{\partial x^{k}}%
\frac{\partial x^{p}}{\partial x^{p^{\prime}}}\frac{\partial x^{q}}{\partial
x^{q^{\prime}}}+\frac{\partial x^{k^{\prime}}}{\partial x^{k}}\frac
{\partial^{2}x^{k}}{\partial x^{p^{\prime}}\partial x^{q^{\prime}}}\right)  ,
\]
see \cite[p. 283]{DFN}.

If we are given a (contravariant) vector field with components $v^{i}(x)$, the
quantities%
\begin{equation}
\widetilde{\partial}_{x^{j}}v^{i}(x):=\partial_{x_{j}}v^{i}(x)+\sum_{k}%
v^{k}(x)\Gamma_{\,k,j}^{i}(x) \label{covder}%
\end{equation}
make up\textsl{ } a $(1,1)-$tensor, called \textsl{covariant derivative} of
$v$ (with respect to the given connection)$.$ Analogously, given a covector
field with components $w_{i}(x),$ the quantities%
\[
\widetilde{\partial}_{x^{j}}w_{i}(x):=\partial_{x^{j}}w_{i}(x)-\sum_{k}%
w_{k}(x)\Gamma_{\,j,i}^{k}(x)
\]
make up a $(2,0)$-tensor, the covariant derivative of $w.$ In a completely
analogous fashion, one can define covariant derivatives of higher rank
tensors, see \cite[p. 281]{DFN}. Covariant derivatives of scalar fields are
just partial derivatives.

It should be remarked that the system of partial derivatives of a vector field
$(\partial_{x_{j}}v^{i}(x))$ does not define a tensor. Covariant derivatives
coincide with usual partial derivatives only in a coordinate system for which
$\Gamma_{\,k,j}^{i}=0$ (so called Euclidean connection). Such Euclidean
connection not always exists.

The operation of covariant differentiation allows to define the parallel
transport of a vector (tensor) field along a curve, as well as the concept of
geodesics (relative to the given connection).

Given a connection, we can define the \textsl{covariant} \textsl{divergence}
of a vector field $v=(v^{i}(x))$ as
\begin{equation}
\widetilde{\operatorname{div}}\,v=\sum_{i}\widetilde{\partial}_{x^{i}}v^{i},
\label{divergence}%
\end{equation}
that is, as the convolution of the $(1,1)-$covariant derivative. By its
definition as a convolution, this quantity is a real (invariant) scalar.

However, if $M$ is a Riemannian manifold, there is a natural way to choose the
connection. Indeed, if $(g_{ij})$ is the metric tensor on $M,$ we require $M$
to be "intrinsically flat" in the sense that%
\begin{equation}
\widetilde{\partial}_{x^{k}}g_{ij}(x)=0\qquad\text{on \ }M. \label{Ricci}%
\end{equation}
The requirement of this property, together with symmetry,%
\[
\Gamma_{\,\,j,k}^{i}=\Gamma_{\,\,k,j}^{i}%
\]
uniquely defines a connection (so called connection \textsl{compatible with
the metrics, or Levi-Civita connection}). The Christoffel symbols are given in
terms of the metrics by the formulas%
\begin{equation}
\Gamma_{\,ij}^{k}=\frac{1}{2}\sum_{l}g^{kl}\left(  \frac{\partial g_{ij}%
}{\partial x^{i}}+\frac{\partial g_{il}}{\partial x^{j}}-\frac{\partial
g_{ij}}{\partial x^{l}}\right)  , \label{levicivita}%
\end{equation}
see \cite[p. 293]{DFN}.

For the Levi-Civita connection, parallel transport preserves lengths and
angles between vectors in their corresponding tangent spaces. Also, the
divergence can be written in terms of the metric tensor as%
\begin{equation}
\widetilde{\operatorname{div}}\,v=\frac{1}{\sqrt{g}}\sum_{i}\frac{\partial
}{\partial x^{i}}\left(  \sqrt{g}v^{i}\right)  =\sum_{i}\frac{\partial v^{i}%
}{\partial x^{i}}+\sum_{i}v^{i}\frac{\partial\log\sqrt{g}}{\partial x^{i}},
\label{divergenceRiemannian}%
\end{equation}
where $g=\det(g_{ij})$. The above formula follows in a straightforward manner
from (\ref{covder}), (\ref{levicivita}) and (\ref{divergence}). Indeed, from
(\ref{divergence}) and (\ref{covder}) we have%
\begin{equation}
\widetilde{\operatorname{div}}\,v=\sum_{i}\widetilde{\partial}_{x^{i}}%
v^{i}=\sum_{i}\partial_{x^{i}}v^{i}+\sum_{i,k}v^{k}\Gamma_{\,k,i}^{i}
\label{eq1}%
\end{equation}
and by (\ref{levicivita}),
\begin{equation}
\sum_{i}\Gamma_{\,k,i}^{i}=\frac{1}{2}\sum_{i,l}g^{il}\left(  \frac{\partial
g_{lk}}{\partial x^{i}}+\frac{\partial g_{il}}{\partial x^{k}}-\frac{\partial
g_{ki}}{\partial x^{l}}\right)  =\frac{1}{2}\sum_{i,l}g^{il}\frac{\partial
g_{il}}{\partial x^{k}}=\frac{1}{2g}\frac{\partial g}{\partial x^{k}}%
=\frac{\partial\log\sqrt{g}}{\partial x^{k}}, \label{eq2}%
\end{equation}
where we have used the symmetry of the metric tensor in the second equality,
and the obvious identity%
\[
\frac{\partial g}{\partial x^{k}}=\sum_{i,l}\frac{\partial g_{il}}{\partial
x^{k}}G_{il}=g\sum_{i,l}\frac{\partial g_{il}}{\partial x^{k}}g_{il},
\]
in the third, where $(G_{il})$ is the matrix of cofactors of $(g_{ij}).$
Plugging (\ref{eq2}) into (\ref{eq1}), we get (\ref{divergenceRiemannian}).

For the compatible connection, the requirement of the existence of an
Euclidean connection is equivalent to the requirement of the manifold being
Euclidean in the usual sense, that is, in the sense that there exists a system
of coordinates for which the metrics is constant along the manifold (this
requirement is clearly equivalent to the existence of coordinates for which
$g_{ij}=\delta_{ij}$ (Cartesian coordinates)). Such coordinates do not exist,
in particular, on a curved submanifold of Euclidean space. In Cartesian
coordinates it follows from (\ref{levicivita}) that $\Gamma_{\,\,j,k}^{i}=0$
and the divergence boils down to the sum of partial derivatives.

\subsubsection{The Divergence theorem\label{DivThm}}

Stokes' Theorem is the main result in the theory of integration of
differential forms. For the Theorem to be valid, no special structure is
required on the manifold. We remind the reader that, according to Stokes'
Theorem, if $\omega$ is a smooth $(k-1)-$differential form defined on a
compact $k-$dimensional manifold $M$ with smooth boundary $\partial M,$ then%
\[
\int_{M}\mathrm{d\omega}=\int_{\partial M}\mathrm{\omega},
\]
where $\mathrm{d\omega}$ stands for the exterior differential of
$\mathrm{\omega,}$ a $k-$form.

If $M$ is an open subset of an $n$-dimensional Riemannian manifold $V^{n}$
with closed boundary $\partial M,$ and both $M,$ $\partial M$ are equipped
with the Riemannian structure of the embedding manifold $V^{n},$ one can
formulate a version of Stokes' Theorem for vector fields, in which the metrics
is displayed explicitly, see \cite[page 283]{LR}. The main reason for this is
that in this case there is a correspondence between $(n-1)$-forms and
contravariant vectors, via the Hodge map and the Riemannian structure. Let
$v^{i}(x)$ be a contravariant vector field on $M$. Thanks to the Riemannian
structure, one can define a properly oriented unit normal vector at each point
of the hypersurface $\partial M$, see \cite{LR}. Let $(u^{i})$ denote some
local coordinates in $\partial M$ and $n=(n^{i}(u^{j}))$ be such a vector.
Then,%
\begin{equation}
\int_{M}\widetilde{\operatorname{div}}\,v\text{ }\mathrm{dV}_{\mathrm{n}}%
=\int_{\partial M}\left\langle v,n\right\rangle \text{ }\mathrm{dV}%
_{\mathrm{n-1}}, \label{StokesRiemannian}%
\end{equation}
where $\widetilde{\operatorname{div}}\,$\ is the divergence defined in
(\ref{divergence}), $\left\langle \,,\right\rangle $ is the scalar product in
$V^{n}$ and%
\[
\mathrm{dV}_{\mathrm{n}}=\sqrt{\det(g_{ij})}\,\mathrm{dx}^{1}\mathrm{\wedge
dx}^{2}\mathrm{\wedge...\wedge dx}^{n};\qquad\mathrm{dV}_{\mathrm{n-1}}%
=\sqrt{\det(\widehat{g}_{ij})}\,\mathrm{du}^{1}\mathrm{\wedge du}%
^{2}\mathrm{\wedge...\wedge du}^{n-1}%
\]
are the volume form in $V^{n}$, respectively the induced volume form in
$\partial M$, in local coordinates $(u^{i}).$

When $V^{n}$ is the Euclidean space $\mathbb{E}^{n}$ with Euclidean
coordinates $(x^{i})$, we recover from (\ref{StokesRiemannian}) the standard
theorems of Vector Calculus. For example, let there a closed, simple, smooth
curve $\gamma:t\in\lbrack a,b]\rightarrow\mathbb{E}^{2}$ be given in the plane
and let us call $M$ its interior. The exterior unit normal vector to $\gamma$
is%
\[
n=\left(  \frac{y^{\prime}(t)}{\sqrt{x^{\prime}(t)^{2}+y^{\prime}(t)^{2}}%
},-\frac{x^{\prime}(t)}{\sqrt{x^{\prime}(t)^{2}+y^{\prime}(t)^{2}}}\right)  ,
\]
where " $^{\prime}$ " indicates derivation with respect to the curve
parameter. \ If a vector field $v(x,y)=(v^{1}(x,y),v^{2}(x,y))$ is given in
$\mathbb{E}^{2},$ as a function of Cartesian coordinates $(x,y),$the
divergence of $v$ is given by the usual formula%
\[
\widetilde{\operatorname{div}}\text{ }v=\frac{\partial v_{1}}{\partial
x}+\frac{\partial v_{2}}{\partial y},
\]
and
\[
\mathrm{dV}=\mathrm{dx\wedge dy;}\qquad\mathrm{dS}=\sqrt{x^{\prime}%
(t)^{2}+y^{\prime}(t)^{2}}\mathrm{dt.}%
\]
Therefore, (\ref{StokesRiemannian}) gives%
\[
\int_{M}\left(  \frac{\partial v_{1}}{\partial x}+\frac{\partial v_{2}%
}{\partial y}\right)  \,\mathrm{dx\wedge dy}=\int_{\gamma}\left(
v^{1}y^{\prime}-v^{2}x^{\prime}\right)  \,\mathrm{dt.}%
\]
Renaming $P=-v^{2},$ $Q=v^{1},$ we can write the previous formula in the
classical form%
\[
\int_{M}\left(  \frac{\partial Q}{\partial x}-\frac{\partial P}{\partial
y}\right)  \,\mathrm{dx\wedge dy}=\int_{\gamma}P\,\mathrm{dx}+Q\,\mathrm{dy}%
\]
(Green's Theorem in the plane).

\subsection{Energy conservation\label{apencon}}

Let us consider a field Lagrangian of the form%
\begin{gather}
L=L\left(  x,\psi,D\psi\right)  ,\quad x=\left(  x^{0},x^{1},\ldots
,x^{n}\right)  ,\quad x^{0}=t,\label{encopsi1}\\
\psi=\left\{  \psi^{\ell},\quad\ell=1,\ldots,k\right\}  ,\nonumber\\
D\psi=\left\{  \partial_{\nu}\psi^{\ell},\quad\nu=0,1,\ldots n,\quad
\ell=1,\ldots,k\right\}  ,\nonumber
\end{gather}
Assuming the existence of external, non-monogenic forces $F^{\ell}$ acting on
the field $\psi^{\ell},$the corresponding Euler-Lagrange equations are,
\begin{equation}%
{\displaystyle\sum\nolimits_{\nu}}
\frac{d}{dx^{\nu}}\frac{\partial L}{\partial\partial_{\nu}\psi^{\ell}}%
-\frac{\partial L}{\partial\psi^{\ell}}=F^{\ell},\quad\ell=1,\ldots,k,
\label{encopsi2}%
\end{equation}
or%
\begin{equation}%
{\displaystyle\sum\nolimits_{\nu}}
\frac{dG^{\ell\nu}}{dx^{\nu}}-\frac{\partial L}{\partial\psi^{\ell}}=F^{\ell
},\quad G^{\ell\nu}=\frac{\partial L}{\partial\partial_{\nu}\psi^{\ell}}%
,\quad\ell=1,\ldots,k. \label{encopsi3}%
\end{equation}
Then the\emph{ canonical (Hamilton) energy-momentum tensor} is defined by%
\begin{equation}
T^{\mu\nu}=%
{\displaystyle\sum\nolimits_{\ell}}
\frac{\partial L}{\partial\partial_{\mu}\psi^{\ell}}\partial^{\nu}\psi^{\ell
}-\delta^{\nu\mu}L=%
{\displaystyle\sum\nolimits_{\ell}}
G^{\ell\nu}\partial^{\nu}\psi^{\ell}-\delta^{\nu\mu}L, \label{encopsi4}%
\end{equation}
and for any solution $\psi^{\ell}$ to the EL equations (\ref{encopsi2}%
)-(\ref{encopsi3}) it satisfies the following conservation equations,
\cite[13.3]{Goldstein}, \cite[3.1]{GiaI}%
\begin{equation}%
{\displaystyle\sum\nolimits_{\mu}}
\partial_{\mu}T^{\mu\nu}=-\frac{\partial L}{\partial x_{\nu}}.
\label{encopsi5}%
\end{equation}
In the case when dissipative forces are present the corresponding terms have
to be added to the right-hand side of the above equation, \cite{FS2}. In
particular, the energy conservation law corresponding to $\nu=0$ in
(\ref{encopsi5}) can be represented as%
\begin{equation}
\frac{dH}{dt}+\operatorname{div}S=-\frac{\partial L}{\partial t}+%
{\displaystyle\sum\nolimits_{\ell}}
F^{\ell}\partial_{0}\psi^{\ell},\quad\operatorname{div}S=\partial_{i}S^{i},
\label{encopsi6}%
\end{equation}
where%
\begin{align}
H  &  =T^{00}=%
{\displaystyle\sum\nolimits_{\ell}}
G^{\ell0}\partial^{0}\psi^{\ell}-L=%
{\displaystyle\sum\nolimits_{\ell}}
\frac{\partial L}{\partial\partial_{t}\psi^{\ell}}\partial_{t}\psi^{\ell
}-L\text{ is the energy density,}\label{encopsi7}\\
S^{i}  &  =T^{0i}=%
{\displaystyle\sum\nolimits_{\ell}}
G^{\ell i}\partial^{0}\psi^{\ell}=%
{\displaystyle\sum\nolimits_{\ell}}
G^{\ell i}\partial_{t}\psi^{\ell},\quad i=1,\ldots n\text{ is the energy
flux.}\nonumber
\end{align}

In fact, the energy conservation law (\ref{encopsi7}) can be derived
straightforwardly as follows, see \cite[Section 4.1.]{Be}. Consider the full
time derivative of the Lagrangian with $\psi$ being a solution to the EL
equations (\ref{encopsi2}):
\begin{gather}
\frac{dL}{dx^{0}}=%
{\displaystyle\sum\nolimits_{\ell}}
\left(  \frac{\partial L}{\partial\psi^{\ell}}\partial_{0}\psi^{\ell}+%
{\displaystyle\sum\nolimits_{\nu}}
\frac{\partial L}{\partial\partial_{\nu}\psi^{\ell}}\partial_{\nu}\partial
_{0}\psi^{\ell}\right)  +\frac{\partial L}{\partial x_{0}}=\label{encopsi8}\\
=%
{\displaystyle\sum\nolimits_{\ell}}
\left[  \frac{\partial L}{\partial\psi^{\ell}}-%
{\displaystyle\sum\nolimits_{\nu}}
\frac{dG^{\ell\nu}}{dx^{\nu}}\right]  \partial_{0}\psi^{\ell}+%
{\displaystyle\sum\nolimits_{\nu}}
\frac{d}{dx^{\nu}}\left(  G^{\ell\nu}\partial_{0}\psi^{\ell}\right)
+\frac{\partial L}{\partial x^{0}}.\nonumber
\end{gather}
Taking into account the EL equations (\ref{encopsi2}) we can recast the
identity (\ref{encopsi8}) as
\begin{equation}
\frac{d}{dx^{0}}\left[
{\displaystyle\sum\nolimits_{\ell}}
G^{\ell0}\partial_{0}\psi^{\ell}-L\right]  +%
{\displaystyle\sum_{i=1}^{n}}
\partial_{i}\left(
{\displaystyle\sum\nolimits_{\ell}}
G^{\ell i}\partial_{0}\psi^{\ell}\right)  =-\partial_{0}L+%
{\displaystyle\sum\nolimits_{\ell}}
F^{\ell}\partial_{0}\psi^{\ell}, \label{encopsi9}%
\end{equation}
which is equivalent to the desired energy conservation (\ref{encopsi6}%
)-(\ref{encopsi7}).

\subsubsection{Covariant Riemann manifold version of the energy conservation}

In the case when the domain $\Theta$ is $n-$dimensional Riemannian manifold
with metric $g_{ij}$ as the boundary $\Gamma$ considered in Section
\ref{SubSectMultipleDimensions} and $L$ is a Lagrangian as in (\ref{encopsi1})
the action $S$ is defined by the following integral
\begin{equation}
S=\int_{t_{0}}^{t_{1}}\mathrm{d}t\int_{\Theta}L\,\mathrm{d}V\mathrm{,}\text{
where }\mathrm{d}V=\sqrt{\det(g_{ij})}\mathrm{d}x^{1}\wedge\mathrm{dx}%
^{2}\wedge\cdots\wedge\mathrm{d}x^{n}. \label{cogcon1}%
\end{equation}
The manifold version of the EL equations (\ref{encopsi2})-(\ref{encopsi3})
reads%
\begin{gather}
\partial_{0}G^{\ell0}+\sum_{i=1}^{n}\tilde{\partial}_{i}G^{\ell i}%
-\frac{\partial L}{\partial\psi^{\ell}}=F^{\ell},\label{cogcon2}\\
G^{\ell\nu}=\frac{\partial L}{\partial\partial_{\nu}\psi^{\ell}},\quad
\nu=0,1,\ldots n,\quad\ell=1,\ldots,k,\nonumber
\end{gather}
where the covariant derivatives $\tilde{\partial}_{i}$ are defined by
(\ref{covder}). In other words, to obtain the manifold version of the EL
equations we simply substitute in the conventional EL equations
(\ref{encopsi2})-(\ref{encopsi3}) the spatial partial derivatives
$\partial_{i}$ with their covariant counterparts $\tilde{\partial}_{i}$. This
follows from the derivation in subsection \ref{DeriMulti}. The same
substitution yields the covariant version of the energy conservation law
(\ref{encopsi6})-(\ref{encopsi7}), namely%
\begin{equation}
\frac{dH}{dt}+\widetilde{\operatorname{div}}\ S=-\frac{\partial L}{\partial
t}+%
{\displaystyle\sum\nolimits_{\ell}}
F^{\ell}\partial_{0}\psi^{\ell},\quad\widetilde{\operatorname{div}}\ S=%
{\displaystyle\sum\nolimits_{i}}
\tilde{\partial}_{i}S^{i}, \label{cogcon4}%
\end{equation}
where%
\begin{align}
H  &  =%
{\displaystyle\sum\nolimits_{\ell}}
G^{\ell0}\partial^{0}\psi^{\ell}-L=%
{\displaystyle\sum\nolimits_{\ell}}
\frac{\partial L}{\partial\partial_{t}\psi^{\ell}}\partial_{t}\psi^{\ell
}-L\text{ is the energy density,}\label{cogcon5}\\
S^{i}  &  =T^{0i}=%
{\displaystyle\sum\nolimits_{\ell}}
G^{\ell i}\partial^{0}\psi^{\ell}=%
{\displaystyle\sum\nolimits_{\ell}}
G^{\ell i}\partial_{t}\psi^{\ell},\quad i=1,\ldots n\text{ is the energy
flux.}\nonumber
\end{align}
To justify the covariant form of the energy conservation law (\ref{cogcon4}%
)-(\ref{cogcon5}) we use the same identity (\ref{encopsi8}) as in the
conventional case and recast it using the explicit formula
(\ref{divergenceRiemannian}). Indeed
\begin{gather}
\frac{dL}{dx^{0}}=%
{\displaystyle\sum\nolimits_{\ell}}
\left[  \frac{\partial L}{\partial\psi^{\ell}}\partial_{0}\psi^{\ell}+%
{\displaystyle\sum\nolimits_{\nu}}
G^{\ell\nu}\partial_{\nu}\partial_{0}\psi^{\ell}\right]  +\frac{\partial
L}{\partial x^{0}}=\label{cogcon6}\\
=%
{\displaystyle\sum\nolimits_{\ell}}
\left[  \frac{\partial L}{\partial\psi^{\ell}}-%
{\displaystyle\sum\nolimits_{\nu}}
\frac{dG^{\ell\nu}}{dx^{\nu}}-\sum_{i=1}^{n}G^{\ell i}\frac{\partial\log
\sqrt{g}}{\partial x^{i}}\right]  \partial_{0}\psi^{\ell}+\nonumber\\
+%
{\displaystyle\sum\nolimits_{\ell}}
\left[
{\displaystyle\sum\nolimits_{\nu}}
\frac{d}{dx^{\nu}}\left(  G^{\ell\nu}\partial_{0}\psi^{\ell}\right)
+\sum_{i=1}^{n}G^{\ell i}\partial_{0}\psi^{\ell}\frac{\partial\log\sqrt{g}%
}{\partial x^{i}}\right]  +\frac{\partial L}{\partial x_{0}}.\nonumber
\end{gather}
Using now the formula (\ref{divergenceRiemannian}) and applying to the
identity (\ref{cogcon6}) the covariant EL equations (\ref{cogcon4}) we obtain
\begin{align}
\frac{dL}{dx^{0}}  &  =%
{\displaystyle\sum\nolimits_{\ell}}
\left[  \frac{\partial L}{\partial\psi^{\ell}}-\frac{dG^{\ell\nu}}{dx^{\nu}%
}-\sum_{i=1}^{n}G^{\ell i}\frac{\partial\log\sqrt{g}}{\partial x^{i}}\right]
\partial_{0}\psi^{\ell}+\frac{\partial L}{\partial x_{0}}-%
{\displaystyle\sum\nolimits_{\ell}}
F^{\ell}\partial_{0}\psi^{\ell}=\label{cogcon7}\\
&  =%
{\displaystyle\sum\nolimits_{\ell}}
\left[  \frac{d}{dx^{0}}\left(  G^{\ell0}\partial_{0}\psi^{\ell}\right)
+\sum_{i=1}^{n}\tilde{\partial}_{i}\left(  G^{\ell i}\partial_{0}\psi^{\ell
}\right)  \right]  +\frac{\partial L}{\partial x_{0}}-%
{\displaystyle\sum\nolimits_{\ell}}
F^{\ell}\partial_{0}\psi^{\ell}.\nonumber
\end{align}
The above equation can be readily recast as%
\begin{equation}
\frac{d}{dx^{0}}\left(
{\displaystyle\sum\nolimits_{\ell}}
G^{\ell0}\partial_{0}\psi^{\ell}-L\right)  +\sum_{i=1}^{n}\tilde{\partial}%
_{i}\left(
{\displaystyle\sum\nolimits_{\ell}}
G^{\ell i}\partial_{0}\psi^{\ell}\right)  =-\frac{\partial L}{\partial x_{0}}+%
{\displaystyle\sum\nolimits_{\ell}}
F^{\ell}\partial_{0}\psi^{\ell}, \label{cogcon8}%
\end{equation}
which is equivalent to the desired covariant form of the energy conservation
law (\ref{cogcon4})-(\ref{cogcon5}).

\end{document}